%% file: HIG-17-014_temp.tex
\pdfoutput=1

\documentclass[11pt,twoside,a4paper,cmspaper,final,collab]{cms-tdr}
\def\svnVersion{4f11b33-D}\def\svnDate{2019/06/22}\def\cmsCernNoTag{CERN-EP-2019-035}\def\cmsCernDate{\today}\def\cmsMessage{"Published in the Journal of High Energy Physics as \href{http://dx.doi.org/10.1007/JHEP05(2019)210}{\doi{10.1007/JHEP05(2019)210}.}"}

\begin{document}\cmsNoteHeader{HIG-17-014}

\hyphenation{had-ron-i-za-tion}
\hyphenation{cal-or-i-me-ter}
\hyphenation{de-vices}
\newcommand{\PNfix}   {\hspace{-.04em}}
\newcommand{\tautau}  {{\ensuremath{\Pgt\PNfix\Pgt}}\xspace}
\newcommand{\mumu}    {{\ensuremath{\Pgm\PNfix\Pgm}}\xspace}
\newcommand{\elel}    {{\ensuremath{\Pe\PNfix\Pe}}\xspace}
\newcommand{\etau}    {{\ensuremath{\Pe\PNfix\Pgt\hspace{-.05em}_\mathrm{h}}}\xspace}
\newcommand{\mutau}   {{\ensuremath{\Pgm\PNfix\Pgt\hspace{-.05em}_\mathrm{h}}}\xspace}
\newcommand{\ltau}    {{\ensuremath{\ell\Pgt\PNfix_\text{h}}}\xspace}
\newcommand{\epsmutau}{{\ensuremath{\epsilon^{\mutau}_\text{reco}}}\xspace}
\newcommand{\epsmumu} {{\ensuremath{\epsilon^{\mumu}_\text{reco}}}\xspace}
\newcommand{\NS}      {{\ensuremath{N_\mathrm{S}}}\xspace}
\newcommand{\X}       {{\ensuremath{\mathrm{X}}}\xspace}
\newcommand{\A}       {{\ensuremath{\PXXA}}\xspace}
\newcommand{\B}       {{\ensuremath{\mathrm{B}}}\xspace}
\newcommand{\bbA}     {{\ensuremath{\bbbar\PNfix\PXXA}}\xspace}
\newcommand{\qbX}     {{\ensuremath{\cPq\cPqb\X}}\xspace}
\newcommand{\mA}      {{\ensuremath{m_{\PXXA}}}\xspace}
\newcommand{\mX}      {{\ensuremath{m_{\X}}}\xspace}
\newcommand{\mB}      {{\ensuremath{m_{\PB}}}\xspace}
\newcommand{\mtt}     {{\ensuremath{m_{\tautau}}}\xspace}
\newcommand{\pzetamiss}{{\ensuremath{p_{\zeta}^\text{miss}}}\xspace}
\newcommand{\pzetavis}{{\ensuremath{p_{\zeta}^\text{vis}}}\xspace}
\newcommand{\SVFit}   {{\textsc{SVfit}}\xspace}
\newcommand{\zjets}   {{\ensuremath{\cPZ\PNfix+\text{jets}}\xspace}}
\newcommand{\wjets}   {{\ensuremath{\PW\PNfix+\text{jets}}\xspace}}

\RCS$Revision: 491727 $
\RCS$HeadURL: svn+ssh://ineuteli@svn.cern.ch/reps/tdr2/papers/HIG-17-014/trunk/HIG-17-014.tex $
\RCS$Id: HIG-17-014.tex 491727 2019-03-13 17:40:53Z ineuteli $
\newlength\cmsFigWidth
\setlength\cmsFigWidth{0.44\textwidth}
\providecommand{\cmsLeft}{left\xspace}
\providecommand{\cmsRight}{right\xspace}
\cmsNoteHeader{HIG-17-014}

\title{Search for a low-mass $\TT$ resonance in association with a bottom quark in proton-proton collisions at $\sqrt{s}=13\TeV$}

\date{\today}

\abstract{
A general search is presented for a low-mass $\TT$ resonance produced in association with a bottom quark. The search is based on proton-proton collision data at a center-of-mass energy of 13\TeV collected by the CMS experiment at the LHC, corresponding to an integrated luminosity of $35.9\fbinv$. The data are consistent with the standard model expectation. Upper limits at 95\% confidence level on the cross section times branching fraction are determined for two signal models: a light pseudoscalar Higgs boson decaying to a pair of $\tau$ leptons produced in association with bottom quarks, and a low-mass boson \X decaying to a {\Pgt}-lepton pair that is produced in the decay of a bottom-like quark \B such that $\B\to\cPqb\X$. Masses between 25 and 70\GeV are probed for the light pseudoscalar boson with upper limits ranging from 250 to 44\unit{pb}. Upper limits from 20 to 0.3\unit{pb} are set on B masses between 170 and 450\GeV for \X boson masses between 20 and 70\GeV. }

\hypersetup{%
pdfauthor={CMS Collaboration},%
pdftitle={Search for a low-mass tau+tau- resonance in association with a bottom quark in proton-proton collisions at sqrt(s)=13 TeV},%
pdfsubject={CMS},%
pdfkeywords={CMS, physics, low-mass, light, pseudoscalar, pseudo-scalar, 2HDM, Higgs doublet, VLQ, vector-like quark, bottom-like, ditau, tau, bottom, b jet, 13 TeV}}

\maketitle

\section{Introduction} \label{sec:introduction}

The observation of a Higgs boson by the ATLAS and the CMS Collaborations~\cite{Higgs_discovery_2012_ATLAS,Higgs_discovery_2012_CMS,Higgs_discovery_2013_CMS} represents a major step towards the understanding of the mechanism for electroweak symmetry breaking~\cite{SM1,Higgs_theory1,Higgs_theory2}.
All measurements within the Higgs boson sector have so far been in general agreement with the predictions of the standard model (SM)~\cite{ATLAS-MASS,HIG-14-009}.
However, the SM cannot address several crucial issues, such as the hierarchy problem, the origin of the matter-antimatter asymmetry in the universe, and the nature of dark matter~\cite{Wess:1974tw,Cheng:2003ju,Appelquist:2000nn,DarkMatt}. Theories beyond the SM have been proposed to address these open questions. Many of these predict the existence of more than one Higgs boson, or new resonances that preferentially decay to a pair of third-generation fermions, including {\Pgt} leptons.

In this analysis, a search for several scenarios of low-mass resonances that decay to a pair of {\Pgt} leptons of opposite charge is performed. In particular, we define multiple signal regions that are optimized based on two benchmark models that have final states with different kinematic properties. We consider a mass range between 20 and 70\GeV, as we are bounded below by our kinematic requirements, and above 70\GeV by the background of the {\cPZ} boson mass peak.

\begin{figure}[b!]
    \centering
    \includegraphics[width=0.32\textwidth]{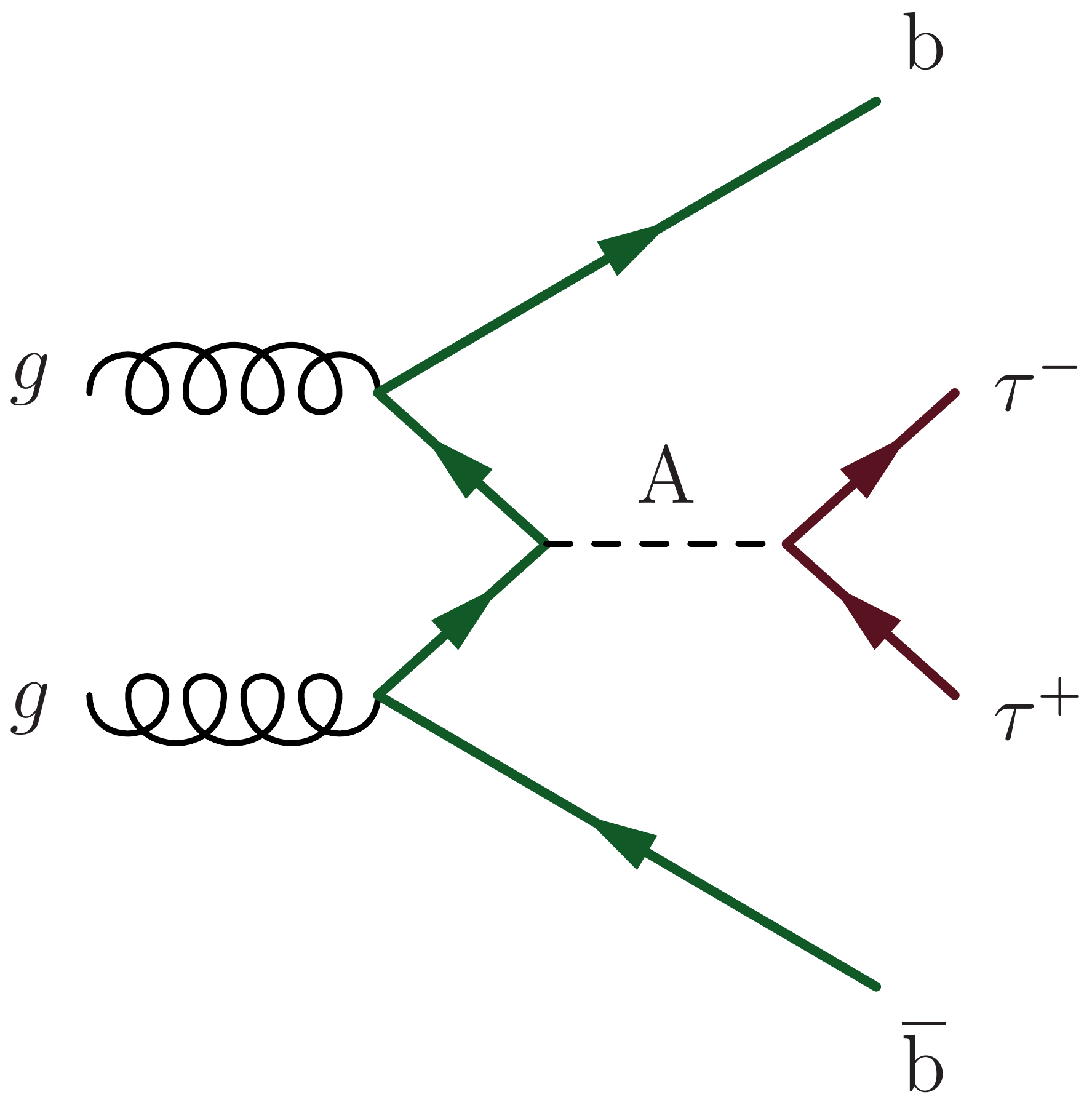}
    \hspace{0.5cm}
    \includegraphics[width=0.37\textwidth]{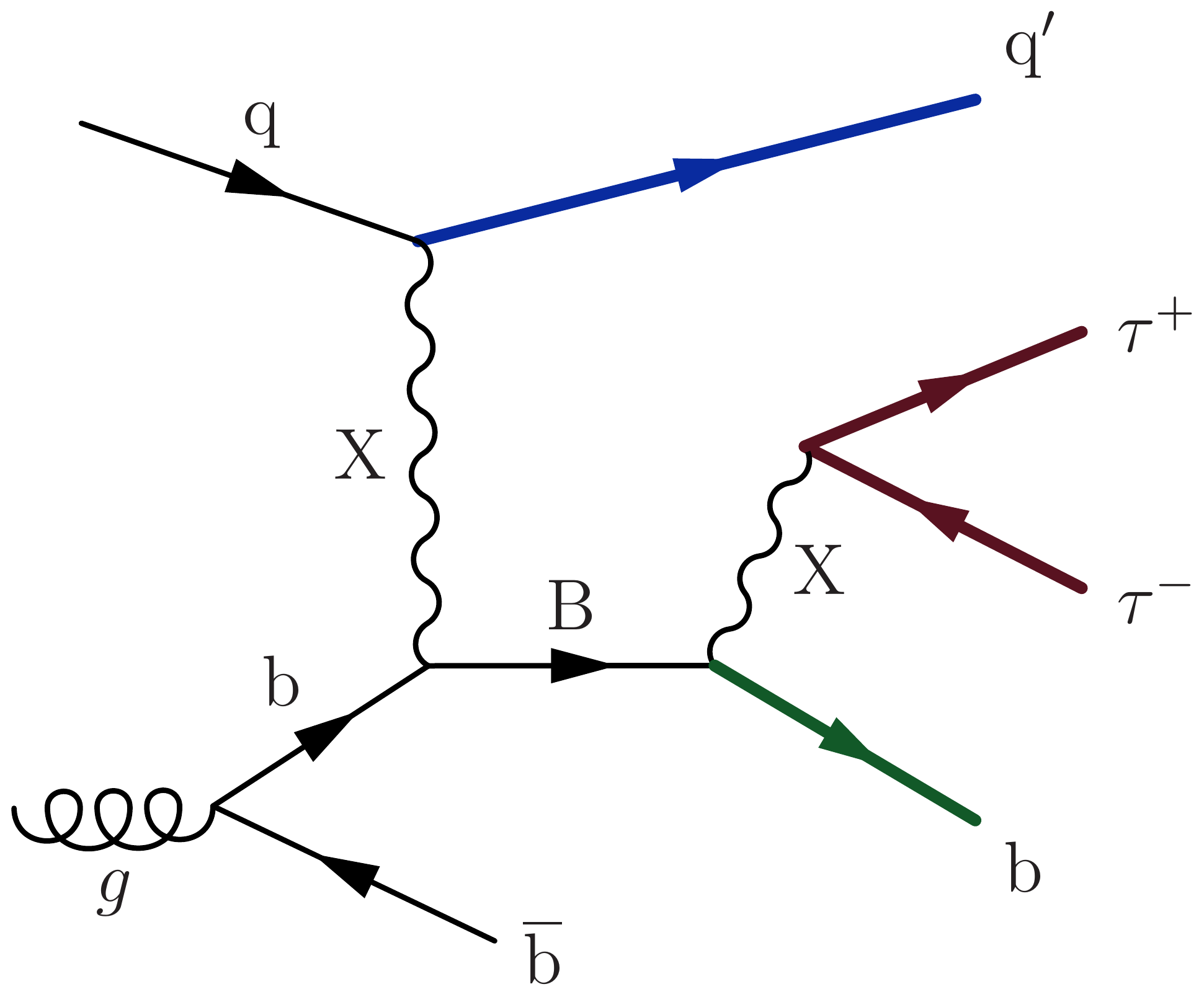}
    \vspace{0.5cm}
    \caption{Feynman diagrams of (\cmsLeft) a low-mass pseudoscalar Higgs boson (A) produced in association with bottom quarks, and (\cmsRight) a bottom-like quark produced in $t$ channel, which decays into X and a bottom quark. The particle \X decays into a {\Pgt}-lepton pair.}
    \label{fig:diagrams}
\end{figure}

The first model describes a low-mass pseudoscalar Higgs boson A, produced in association with two bottom quarks (\bbA), and decaying to a {\Pgt}-lepton pair. This is one of the preferred scenarios in the Two-Higgs-Doublet Models (2HDMs)~\cite{2HDM1,2HDM2,2HDM3,2HDM4,2HDM5}.
Searches for signatures of \bbA or A pair production containing {\Pgt} leptons in the final state have been performed using pp collision data at a center-of-mass energy of 8\TeV collected by CMS~\cite{pseudoscalar_8TeV,pseudoscalar_8TeV_all} and ATLAS~\cite{pseudoscalar_8TeV_ATLAS_mumutautau}, as well as with data at 13\TeV by CMS~\cite{pseudoscalar_13TeV_mumutautau,pseudoscalar_13TeV}.
Other searches by CMS and ATLAS for low-mass bosons exploit final states containing muons and b quarks~\cite{pseudoscalar_8TeV_CMS_bbmumu,pseudoscalar_13TeV_ATLAS_bbmumu,pseudoscalar_13TeV_ATLAS_4b}, but also electrons~\cite{pseudoscalar_8TeV_ATLAS_4l,pseudoscalar_13TeV_ATLAS_4l} or photons~\cite{pseudoscalar_8TeV_ATLAS_4g}.
For this model, we choose events with a {\Pgt}-lepton pair and a central jet that is consistent with the decay of a {\cPqb} hadron (``{\cPqb}-tagged jet''). A Feynman diagram of this signal process at leading order (LO) is shown in Fig.~\ref{fig:diagrams} (left panel).

The second model describes a low-mass boson X decaying to a {\Pgt}-lepton pair in a process where the \X boson is created through the decay of a vector-like quark (VLQ)~\cite{VLQ1,VLQ2,VLQ_LHC1,VLQ_LHC2}.
In the scenario considered here, a heavy bottom-like quark \B is produced in a $t$-channel process
in association with a light quark, where an \X boson acts as the propagator. It then decays via $\B\to\cPqb\X$, so that the final state topology is \qbX. The \B is typically scattered in the forward direction, and two categories of event selection are optimized to target this signature.
Both categories require a jet consistent with the decay of a {\cPqb} hadron, with one category requiring an additional central jet with pseudorapidity $\abs{\eta} < 2.4$, and one category requiring an additional forward jet with $\abs{\eta} > 2.4$. With this selection, the analysis provides new sensitivity to vector-like quarks by targeting previously unexplored decays of heavy bottom-like quarks. The Feynman diagram of this signal process that is dominant at LO is also shown in Fig.~\ref{fig:diagrams} (right panel).

A number of other scenarios beyond the SM produce signatures similar to the two models considered. For example, Hidden Valley models~\cite{HVM1,HVM2} predict a spin-one resonance decaying to lepton pairs; dark-force models~\cite{DFM1} include the decay of a top quark to a bottom quark and two GeV-scale bosons, $\PWpr$ and $\PZpr$, that decay to leptons~\cite{lowmassZ2,Dalchenko:2017shg}; and new flavor changing neutral current interactions of the top quark, in which a new light \X boson is produced in association with a single top quark and decays to lepton pairs~\cite{FCNC1}. Although these new physics scenarios are not considered in this analysis, the results can be applied to most of these cases in the kinematic regions explored in this work.

A previous analysis of proton-proton ($\Pp\Pp$) collision data taken at a center-of-mass energy of 8\TeV, exploring a similar final state focusing on dimuon resonances, has observed excesses at an invariant mass of 28\GeV that correspond to local significances of 4.2 and 2.9 standard deviations in the two event categories defined by the analysis~\cite{dimuon_CMS}.
Reference~\cite{dimuon_CMS} also reports an analysis of data with a center-of-mass energy of 13\TeV, and finds both a $2.0$ standard deviation excess and a 1.4 standard deviation deficit in the same two event categories, respectively. If there were a new heavy particle that had Yukawa-like couplings proportional to mass, the rate would be enhanced in the $\tautau$ final state considered in this work, and would provide additional information on the couplings of such a new particle. Therefore, the results of this analysis are compared to those of Ref.~\cite{dimuon_CMS}.

This analysis is based on $\Pp\Pp$ collision data delivered by the LHC at CERN at a center-of-mass energy of 13\TeV. The data set corresponds to an integrated luminosity of $35.9\fbinv$, collected by the CMS detector during 2016. Only the semileptonic final states $\etau$ and $\mutau$ are considered, where one of the {\Pgt} leptons decays into light leptons (electron or muon), and the other decays hadronically, denoted as $\tauh$.

\section{The CMS detector} \label{sec:detector}

The central feature of the CMS apparatus is a superconducting solenoid of 6\unit{m} internal diameter, providing a magnetic field of 3.8\unit{T}. Within the solenoid volume, there are a silicon pixel and strip tracker, a lead tungstate crystal electromagnetic calorimeter (ECAL), and a brass and scintillator hadron calorimeter, each composed of a barrel and two endcap sections. Forward calorimeters extend the pseudorapidity coverage provided by the barrel and endcap detectors from $\abs{\eta}<3.0$ to $\abs{\eta}<5.2$. Muons are measured in gas-ionization detectors embedded in the steel flux-return yoke outside the solenoid.

Events of interest are selected using a two-tiered trigger system~\cite{trigger}. The first level, composed of custom hardware processors, uses information from the calorimeters and muon detectors to select events at a rate of around 100\unit{kHz} within a time interval of less than 4\mus. The second level, known as the high-level trigger, consists of a farm of processors running a version of the full event reconstruction software optimized for fast processing, and reduces the event rate to about 1\unit{kHz} before data storage.

A more detailed description of the CMS detector, together with a definition of the coordinate system used and the relevant kinematic variables, can be found in Ref.~\cite{CMS}.

\section{Simulated samples} \label{sec:samples}

Samples of simulated events are used to devise selection criteria, and estimate and validate background predictions.
The main sources of background are the pair production of top quarks ($\ttbar$), single top quark production, $\PW$ and $\cPZ$ boson production in association with jets, denoted as ``$\wjets$'' and ``$\zjets$'', diboson ($\PW\PW$, $\PW\cPZ$, $\cPZ\cPZ$) production, and quantum chromodynamics (QCD) production of multijet events.
The $\wjets$ and $\zjets$ processes are simulated using the \MGvATNLO~\cite{aMCatNLO} generator~(2.2.2 and 2.3.3) at LO precision with the MLM jet matching and merging scheme~\cite{MLM}.
The same generator is also used for diboson production simulated at next-to-leading order (NLO) precision with the FxFx jet matching and merging scheme~\cite{FxFx}, whereas \POWHEG~\cite{POWHEG0,POWHEG1,POWHEG2} 2.0 and 1.0 are used for $\ttbar$ and single top quark production at NLO precision, respectively~\cite{POWHEG_TT,POWHEG_TT2,POWHEG_ST_st,POWHEG_ST_tW}.
The $\zjets$, $\ttbar$, and single top processes are normalized using cross sections computed at next-to-next-to-leading order (NNLO) in perturbative QCD~\cite{sigma_DY,sigma_TT,sigma_ST}.

The $\bbA$ samples are produced with the \PYTHIA 8.212~\cite{Pythia8_2015} generator with the pseudoscalar mass ($\mA$) ranging from 25 to 70\GeV.

The \qbX signals are generated with \MGvATNLO, using the same production mechanism as for producing single top quarks in the $t$-channel.
The {\cPqb} quark that initiates the \qbX process is predominantly produced in gluon splittings, and is modeled by the four-flavor scheme (4FS), such that the {\cPqb} quark is not contained in the proton parton distribution functions.
A previous comparison with data has shown that the absolute value of the transverse momentum ($\pt = \abs{\ptvec}$) and $\eta$ distributions of the top quark in simulated $t$-channel events is better modeled in the 4FS than in the five-flavor scheme~\cite{CMS-TOP-17-011}.
Several samples with different values of $\mX$, ranging from 20 to 70\GeV, are generated. Mass values of 170, 300, and 450\GeV are considered for the $\B$ particle.

The event generators are interfaced with \PYTHIA to model the parton showering and fragmentation, as well as the decay of the {\Pgt} leptons. The \PYTHIA parameters affecting the description of the underlying event are set to the {CUETP8M1} tune~\cite{CUETP8M1}. The NNPDF3.0 parton distribution functions~\cite{NNPDF30} with the order matching that of the matrix element calculations are used with all generators.
Generated events are processed through a simulation of the CMS detector based on \GEANTfour~\cite{GEANT4}, and are reconstructed with the same algorithms used for data.
The simulated samples include additional $\Pp\Pp$ interactions per bunch crossing, referred to as ``pileup''.
The effect of pileup is taken into account by generating concurrent total inelastic collision events with \PYTHIA. The simulated events are weighted such that the distribution of the number of pileup interactions matches that in data, with an average of approximately 23 interactions per bunch crossing~\cite{CMS-PAS-LUM-17-001}.

\section{Event and object reconstruction} \label{sec:object}

The reconstruction of observed and simulated events relies on the particle-flow (PF) algorithm~\cite{PF2017}, which combines information from the CMS subdetectors to reconstruct and identify the particles emerging from the $\Pp\Pp$ collisions: charged and neutral hadrons, photons, muons, and electrons.
This section describes how these PF objects are combined to reconstruct other physics objects such as jets, $\tauh$ candidates, or missing transverse momentum ($\ptvecmiss$). The primary $\Pp\Pp$ interaction vertex of an event is taken to be the reconstructed vertex with the largest value of summed physics-object $\PT^2$.

After being reconstructed by the PF algorithm, electrons are identified with a multivariate analysis (MVA)~\cite{TMVA} discriminant that combines several quantities describing the track quality, the shape of the energy deposits in the ECAL, and the compatibility of the measurements from the tracker and the ECAL~\cite{electron}.
Selected electrons must pass a discriminant requirement that rejects electrons coming from photon conversions.
Muons are identified with requirements on the quality of the track reconstruction and on the number of measurements in the tracker and the muon system~\cite{Sirunyan:2018fpa}.
To reject nonprompt or misidentified leptons, a relative lepton isolation $I_{\ell}$ ($\ell = \Pe, \Pgm$) is defined as follows:
\begin{equation*}
I_{\ell} \equiv \frac{\sum_\text{charged} \PT + \max\left( 0, \sum_\text{neutral} \PT
                                         - \frac{1}{2} \sum_\text{charged, PU} \PT \right )}{\PT^{\ell}}.
\end{equation*}
In this expression, $\sum_\text{charged} \PT$ is the scalar \pt sum of the charged hadrons originating from the primary vertex, and located in a cone of size $\Delta R =0.3$ (0.4) centered on the electron (muon) direction, where $\DR = \sqrt{\smash[b]{(\Delta \eta)^2 + (\Delta \phi)^2}}$, $\Delta\eta$ is the difference in pseudorapidity, and $\Delta\phi$ is the difference in azimuthal angle in radians. The sum $\sum_\text{neutral} \PT$ represents the same quantity for neutral hadrons and photons.
The contribution of pileup photons and neutral hadrons is estimated from the scalar \pt sum of charged hadrons originating from pileup vertices, $\sum_\text{charged, PU} \PT$. This sum is multiplied by a factor of $1/2$, which corresponds approximately to the ratio of neutral- to charged-hadron production in the hadronization process of inelastic pp collisions, as estimated from simulation.
In this analysis, $I_{\Pe} < 0.10$ ($I_{\Pgm} < 0.15$) is used as the isolation requirement for the electron (muon).

Jets are reconstructed from PF candidates using the anti-\kt clustering algorithm with a distance parameter of 0.4, implemented in the \FASTJET library~\cite{antikT,fastjet1,fastjet2}.
Charged PF candidates not associated with the primary vertex of the interaction are not considered when reconstructing jets.
An offset correction is applied to jet energies to take into account the contribution from additional $\Pp\Pp$ interactions within the same or nearby bunch crossings~\cite{CMS-JME-10-011}. The energy of a jet is calibrated based on simulation and data through correction factors~\cite{CMS-JME-10-011}. Further identification requirements are applied to distinguish genuine jets from those arising from pileup~\cite{jetPUID}, and additional selection criteria on the energy fractions and multiplicity of charged and neutral particles are applied to each event to remove spurious jet-like features originating from isolated noise patterns in certain HCAL regions~\cite{jetID_2016}.
In this analysis, jets are required to have $\pt>30\GeV$ and $\abs{\eta}<4.7$, and must be separated from the selected leptons by $\Delta R>0.5$.
Jets originating from the hadronization of bottom quarks are identified using the combined secondary vertex algorithm~\cite{btagging}, which exploits observables related to the long lifetime and large mass of {\cPqb} hadrons.
The chosen {\cPqb}-tagging working point corresponds to an identification efficiency of approximately 60\% with
a misidentification rate of approximately 1\% for jets originating from light quarks or gluons, and about 13\% for jets originating from charm quarks.

The $\tauh$ candidates are reconstructed with the hadron-plus-strips algorithm~\cite{HPS2}, which is seeded with anti-\kt jets.
This algorithm reconstructs $\tauh$ candidates based on the number of charged hadrons and on the number of strips of ECAL crystals with energy deposits in the one-prong, one-prong + $\Pgpz$, and three-prong decay modes.
An MVA-based discriminant, including the isolation and lifetime information, is used to reduce the incidence of jets being misidentified as $\tauh$ candidates. The typical working point of this MVA-based isolation discriminant, as used in this analysis, has an efficiency of about 60\% for a genuine $\tauh$, with about a 0.1\% misidentification rate for quark and gluon jets.
Electrons and muons misidentified as $\tauh$ candidates are suppressed using dedicated criteria
based on the consistency between the measurements in the tracker, calorimeters, and muon system.

The vector $\ptvecmiss$ is defined as the negative vectorial sum of the $\ptvec$ of all PF candidates~\cite{CMS-PAS-JME-16-004,CMS-PAS-JME-17-001} originating from the primary vertex. The $\ptvecmiss$ is adjusted for the effect of jet energy corrections.
Recoil corrections are applied to account for the mismodeling of $\ptvecmiss$ in simulated events of the $\zjets$ and $\wjets$ processes.
The corrections are performed on the variable that is defined as the vectorial difference between the measured $\ptvecmiss$ and the total $\ptvec$ of neutrinos originating from the decay of the $\PW$ or $\cPZ$ boson. On average, this reduces the $\ptvecmiss$ obtained from simulation by a few GeV.

\section{Event selection} \label{sec:selection}

The search is performed in events containing \etau or \mutau (collectively \ltau) candidates, produced in association with a {\cPqb}-tagged jet.

In order to select the \etau (\mutau) final states of the {\Pgt}-lepton pair, the trigger requirements are at least one isolated electron (muon) with $\pt>25$ (22)\GeV, or the combination of at least one isolated electron (muon) with $\pt>24$ (19)\GeV and one $\tauh$ candidate with $\pt>20\GeV$.
In addition to the trigger requirements, a common ``baseline selection'' is applied, requiring the events to be consistent with the \ltau signature. Additional event selections to target the \bbA and \qbX signatures are described in the following sections.

\subsection{Baseline selection}

The $\etau$ channel requires one electron candidate with $\PT > 25\GeV$, $\abs{\eta}<2.1$,
and relative isolation (defined in Section~\ref{sec:object}) less than 0.10. The electron should be within a longitudinal distance $d_{z}$ of $0.2 \cm$ and a radial distance $d_{xy}$ of $0.045\cm$ with respect to the primary vertex. One \tauh candidate is required to have $\PT>20\GeV$, $\abs{\eta}<2.3$, and to pass the working point of the MVA-based isolation, as detailed in Section~\ref{sec:object}. The selected electron and $\tauh$ should have an opening angle of $\Delta R > 0.5$ and have opposite-sign (OS) electric charges.
If multiple $\tauh$ candidates are found, the one with the best MVA-based isolation is selected.

Similarly, $\mutau$ events are selected by requiring one muon candidate with $\PT > 20\GeV$ and $\abs{\eta}<2.1$. The relative isolation is taken to be less than $0.15$. The same $d_{z}$ and $d_{xy}$ requirements as those imposed on electron candidates are applied to muons.
The $\tauh$-candidate selection is the same as for \etau events.

For both the \etau and \mutau channels, events with additional isolated electrons (or muons) with $\pt > 10\GeV$ and $\abs{\eta} < 2.5$ (2.1) that pass the same $d_{z}$ and $d_{xy}$ requirements, but a looser identification requirement, are discarded to reduce $\zjets$, \ttbar production, and diboson backgrounds, as well as to keep orthogonality between the $\etau$ and $\mutau$ channels.

\subsection{Additional selection for the \texorpdfstring{$\bbA$}{bbA} search} \label{sec:categories_bbA}
Signal events of the \bbA process are characterized by a {\Pgt}-lepton pair and two bottom quarks.
In order to increase the signal purity, candidate events are required to have at least one {\cPqb}-tagged jet with $\pt>30\GeV$ and $\abs{\eta} < 2.4$.
To further remove $\ttbar$ background, events are required to have a transverse mass ($\mT$) less than 40\GeV, where $\mT$ is defined as
\begin{equation*}
  \mT = \sqrt{{2 \pt^\ell \abs{\ptvecmiss} (1-\cos \Delta\phi)}},
\end{equation*}
in which $\pt^\ell$ is the \PT of the lepton and $\Delta\phi$ is the azimuthal angle between the lepton direction and the \ptvecmiss vector, which here is assumed to be due to the momenta of undetected neutrinos.

In addition, events are required to satisfy \mbox{$\pzetamiss - 0.85 \pzetavis > -40\GeV$}, where $\pzetamiss$ is the component of the $\ptvecmiss$ along the bisector of the \ptvec of the lepton and $\tauh$, while $\pzetavis$ is the sum of the parallel components of the lepton and $\tauh$-candidate \ptvec~\cite{pzeta_CDF}.
This variable quantifies the compatibility of events with the topology wherein the direction of neutrinos from the {\Pgt}-lepton decays are aligned with the direction of the visible {\Pgt}-lepton decay products. This requirement is optimized to remove a substantial amount of \ttbar as well as \wjets\ events.

\subsection{Additional selection for the \texorpdfstring{$\qbX$}{qbX} search} \label{sec:categories}
The final-state bottom quark from $\cPq\cPqb \to \cPq'\B \to \cPq'\cPqb\X$ tends to be more centrally produced with a hard $\PT$ spectrum, whereas the final-state light quark tends to be more forwardly scattered.
This motivates two mutually exclusive categories of events. The first category requires one forward jet and one {\cPqb}-tagged jet, and is labeled as ``1b1f''. Namely,
\begin{itemize} \label{item:category1}
    \item          one {\cPqb}-tagged jet  with $\PT>30\GeV$ and $\abs{\eta}<2.4$;
    \item at least one forward   jet  with $\PT>30\GeV$ and $2.4<\abs{\eta}<4.7$;
    \item     no other           jets with $\PT>30\GeV$ and     $\abs{\eta}<2.4$.
\end{itemize}
The second category, labeled as ``1b1c'', has only two central jets:
\begin{itemize} \label{item:category2}
    \item         one {\cPqb}-tagged jet  with $\PT>30\GeV$,    $\abs{\eta}<2.4$;
    \item exactly one other central jet  with $\PT>30\GeV$ and $\abs{\eta}<2.4$;
    \item         no forward jets with $\PT>30\GeV$ and $2.4<\abs{\eta}<4.7$.
\end{itemize}
In order to further reduce the dominant $\ttbar$ background, an additional requirement of $\mT < 60\GeV$ is applied to events in both categories.
This selection helps to reduce the $\ttbar$ background by a factor of five in 1b1f, and by a factor of two in the 1b1c category, while maintaining a signal acceptance of 91 and 98\%, respectively. Of all selected data events, 18\% fall into 1b1f, and 82\% into 1b1c.

After applying the event selection, an excess of events over the SM backgrounds is searched for using the distribution of the invariant mass of the {\Pgt}-lepton pair, constructed using the \SVFit mass algorithm~\cite{SVFit2014,SVFit2017}.
This algorithm approximates the invariant mass of the $\tautau$ system by exploiting information on the four-vectors of the lepton and $\tauh$, combined with the $xy$-components of \ptvecmiss and its covariance matrix.
For better energy resolution, the $\tauh$ decay modes (one-prong, one-prong + $\Pgpz$, and three-prong) are treated separately.
Although the visible mass of the lepton and $\tauh$ system, defined as the invariant mass of the sum of four-vector from the visible particles, can be also used as a discriminant, the \SVFit mass $\mtt$ is preferred since its peak position locates the resonance mass, while performing equally well in terms of the expected sensitivity.
Considering that the typical resolution of the $\mtt$ distribution is 10--15\%~\cite{SVFit2014,SVFit2017}, a bin width of 5\GeV is chosen.
The maximum likelihood fit method~\cite{LHC-HCG-Report} is performed for the signal extraction, as detailed in Section~\ref{sec:results}.

\section{Background estimation} \label{sec:background}

The dominant background in all search channels and categories comes from $\ttbar$ production because of the presence of genuine electrons, muons, $\tau$ leptons, and bottom quark jets from $\ttbar$ decays.
At lower masses, the QCD multijet background also becomes relevant, while around 90\GeV, there is a considerable $\zjets$ contribution.
Additional small backgrounds are $\wjets$, diboson, and single top quark events.

For the $\bbA$ search, simulated events are used to model $\ttbar$ backgrounds, both for the normalization and the shape of the \SVFit mass distribution.
The normalization of the $\ttbar$ background is checked by defining a control region with a high $\ttbar$ purity and little signal contamination by requiring $\abs{\ptvecmiss} > 60\GeV$ and $\mT > 60\GeV$. All other selection requirements stay the same. The data and simulation show close agreement within statistical uncertainty. Therefore, simulated events are used to predict the yield of $\ttbar$ background processes in the signal region without scaling, as well as the associated uncertainties in the cross section.

For the \qbX search, on the other hand, additional requirements on the jet multiplicity can cause mismodeling of the $\ttbar$ background. A control region is defined with the same jet category selections as described in Section~\ref{sec:categories}, as well as $\abs{\ptvecmiss} > 60\GeV$ and $\mT > 60\GeV$ requirements. The data-to-simulation scale factors for the $\ttbar$ events are then calculated such that the simulated number of events agrees with data in these sidebands. In the $\etau$ ($\mutau$) channel, the scale factor is found to be 0.82\,(0.85) for the 1b1f category, and 1.02\,(0.97) for the 1b1c category. The statistical uncertainties in these scale factors are up to 6\% and considered as nuisance parameters in the combined fit.

The QCD multijet background, in which one jet is misidentified as a $\tauh$ candidate and another as a lepton, is small and is estimated using a control region where the lepton and the $\tauh$ candidate have same-sign (SS) electric charges. In this control region, the QCD multijet yield is obtained by subtracting from the data the contribution from the $\zjets$, $\ttbar$, $\wjets$, and other SM background processes, as determined from simulation.
The expected contribution of the QCD multijet background in the OS signal region is then derived by rescaling the yield obtained in the SS control region by a factor of 1.1, which is measured using a high-purity QCD multijet
sample obtained by inverting the lepton isolation requirement.
The QCD multijet background estimation results in up to 20\% rate uncertainties, accounting for the statistical precision in the region where the extrapolation factor from the SS to OS region is measured. This uncertainty also covers potential dependencies of the OS/SS extrapolation factors on the invariant $\tautau$ mass.

For the $\wjets$ background, the shape is modeled on the basis of simulated events, while its normalization is determined from data using a sideband with $\mT >80\GeV$.
The $\wjets$ simulation is normalized such that the overall yield of the simulated events, including the QCD contribution estimated above, matches the data yield in the sideband with $\mT > 80\GeV$ after the baseline selection but before any jet selection.
The scale factor necessary for the $\wjets$ simulated events is found to be 0.95.
The uncertainties in the $\wjets$ event yields estimated from data are as large as 5\%.
This uncertainty accounts for the statistical limitation of data in the high-$\mT$ sideband, the statistical limitation of the simulated $\wjets$ sample, the systematic uncertainties of other processes in the same region, and the extrapolation from high- to low-$\mT$ regions.

Minor backgrounds, such as diboson and single top quark processes, are estimated from simulation.

\section{Systematic uncertainties} \label{sec:systematics}

A binned maximum likelihood fit of the observed \mtt distribution is used to search for a possible signal over the expected background. The \mtt range from 0 to 350\GeV is used, such that the backgrounds can be constrained by data in the high mass sideband, where the signal is not expected.

Systematic uncertainties may affect the normalization or the shape of the \mtt distribution of the signal and background processes.
These uncertainties are represented by nuisance parameters in the fit, as described below, and summarized in Table~\ref{tab:nuisance}.
We note that systematic uncertainties play a small role in this analysis, as the measurement is ultimately limited by the size of the data sample.

\subsection{Normalization uncertainties}

The uncertainty in the integrated luminosity amounts to 2.5\%~\cite{CMS-PAS-LUM-17-001} and affects the normalization of the signal and background processes that are based on simulation.
Uncertainties in the electron or muon identification and trigger efficiency amount to 2\% each~\cite{HTT_2016_paper}.
The $\tauh$ identification and trigger efficiency have been measured using the ``tag-and-probe'' technique~\cite{HPS2} and an overall rate uncertainty of 10\% is assigned.
For events where electrons or muons are misidentified as $\tauh$ candidates, predominantly $\cPZ\to\elel$ events in the $\etau$ channel and $\cPZ\to\mumu$ events in the $\mutau$ channel, a rate uncertainties of 12 and 25\%~\cite{HPS3}, respectively, are applied, as determined by a tag-and-probe method.
The acceptance uncertainty because of the {\cPqb} tagging efficiency (mistag rate) has been determined to be 3 (5)\%.
The momentum scale uncertainty in $\abs{\ptvecmiss}$~\cite{CMS-PAS-JME-16-004,CMS-PAS-JME-17-001} affects the event yields due to selection requirements on the $\mT$ variable and is estimated to be up to $4\%$.
The uncertainties in the $\wjets$ event yields estimated from data can be as large as $5\%$, as detailed in Section~\ref{sec:background}.
The QCD multijet background estimation is found to have rate uncertainties up to $20\%$.
The normalization uncertainty on the $\zjets$ yield is estimated using a dedicated control region in events with two $\tauh$ candidates and at least one {\cPqb}-tagged jet. A 20\% uncertainty is assigned to the $\zjets$ normalization on the basis of the expected fluctuations in the total number of data events in this control region.
For the $\ttbar$ background, an uncertainty of 6\% in the cross section is computed for the 1 {\cPqb} tag category~\cite{sigma_TT}, while in the 1b1f and 1b1c categories, a 6\% uncertainty is determined from a control region, as previously described.
The uncertainties in the cross section for the diboson and single top quark processes are $6$ and $5.5\%$, respectively.

Finally, theoretical uncertainties in the \bbA cross section calculation due to NNLO corrections
for A masses below 50\GeV increase significantly, as is shown in Fig.~263 of Ref.~\cite{deFlorian:2016spz}.
Therefore, a conservatively estimated uncertainty of 50\% is assigned to the \bbA signal yield.

\subsection{Shape uncertainties}

The stability of the shape and the normalization of the \mtt distribution are tested with respect to the uncertainties in the $\tauh$ and jet energy scales for the signal and background processes.
The uncertainty is estimated by varying the $\tauh$ and jet energies within their respective uncertainties and recomputing \mtt after the final selection.
The uncertainty in the $\tauh$ energy scale amounts to 3\%~\cite{HPS2}, and the uncertainties in the jet energy scale are up to 4\%, depending on the jet $\pt$ and $\eta$~\cite{CMS-JME-10-011}.
However, the variation of the $\mtt$ distribution due to the jet energy scale is found to be negligible, and therefore, only normalization uncertainties of 4\% are considered.
Similarly, for events where a jet, muon, or electron is misidentified as a $\tauh$ candidate, a shape uncertainty is derived by varying the reconstructed $\pt$ of the $\tauh$ candidate by 3\%, and recomputing \mtt after the final selection. The variations due to the electron and muon momentum scales are found to be negligible.

Finally, uncertainties related to the limited number of simulated events are taken into account. They are considered for all bins of the distributions that are used to extract the results. They are uncorrelated across the different samples and across the bins of a single distribution.

\begin{table}[p]
\centering
\topcaption{Sources of systematic uncertainties and their effects on the acceptance or shape resulting from a variation of the nuisance parameter equivalent to one standard deviation.}
\begin{tabular}{lcrr} \hline
  Systematic source               & Involved processes         & \multicolumn{2}{c}{Change in acceptance or shape}  \\
  & & $\etau$ & $\mutau$ \\
\hline
  Integrated luminosity           & Simulated processes        & \multicolumn{2}{c}{2.5\%} \\
  Electron ident. \& trigger      & Simulated processes        & 2\% & \NA \\
  Muon ident. \& trigger          & Simulated processes        & \NA & 2\% \\
  \tauh ident. \& trigger         & Simulated processes        & \multicolumn{2}{c}{10\%}\\
  e misidentified as $\tauh$      & $\cPZ\to\elel$             & 12\% & \NA \\
  $\Pgm$ misidentified as $\tauh$  & $\cPZ\to\mumu$            & \NA & 25\%  \\
  {\cPqb} tagging efficiency, mistag rate & Simulated processes & \multicolumn{2}{c}{3--5\%} \\
  $|\ptvecmiss|$ scale            & Simulated processes        & \multicolumn{2}{c}{Up to 4\%} \\
  $\wjets$ normalization          & $\wjets$                   & \multicolumn{2}{c}{5\%} \\
  QCD multijet normalization      & QCD multijet               & \multicolumn{2}{c}{20\%}\\
  $\zjets$ normalization          & $\cPZ\to\tautau$           & \multicolumn{2}{c}{20\%}\\
  $\ttbar$ normalization          & $\ttbar$ (1b1f, 1b1c only) & \multicolumn{2}{c}{6\%} \\
  $\ttbar$ cross section          & $\ttbar$ (\bbA only)       & \multicolumn{2}{c}{6\%} \\
  Diboson cross section           & Diboson & \multicolumn{2}{c}{6\%} \\
  Single top quark cross section  & Single top quark           & \multicolumn{2}{c}{5.5\%} \\
  $\bbA$ cross section            & Signal (\bbA only)         & \multicolumn{2}{c}{50\%} \\
  $\tauh$ energy scale            & Simulated processes        & \multicolumn{2}{c}{Shape} \\
  $\Pe/\Pgm \to \tauh$ energy scale & Simulated processes      & \multicolumn{2}{c}{Shape} \\
  Jet energy scale                & Simulated processes        & \multicolumn{2}{c}{4\%} \\
  Jet misidentified as $\tauh$    & $\zjets$                   & \multicolumn{2}{c}{Shape} \\
  Limited event count             & All processes              & \multicolumn{2}{c}{Shape} \\
\hline
\end{tabular}
\label{tab:nuisance}
\end{table}

\begin{figure}[p]
  \centering
  \includegraphics[width=\cmsFigWidth]{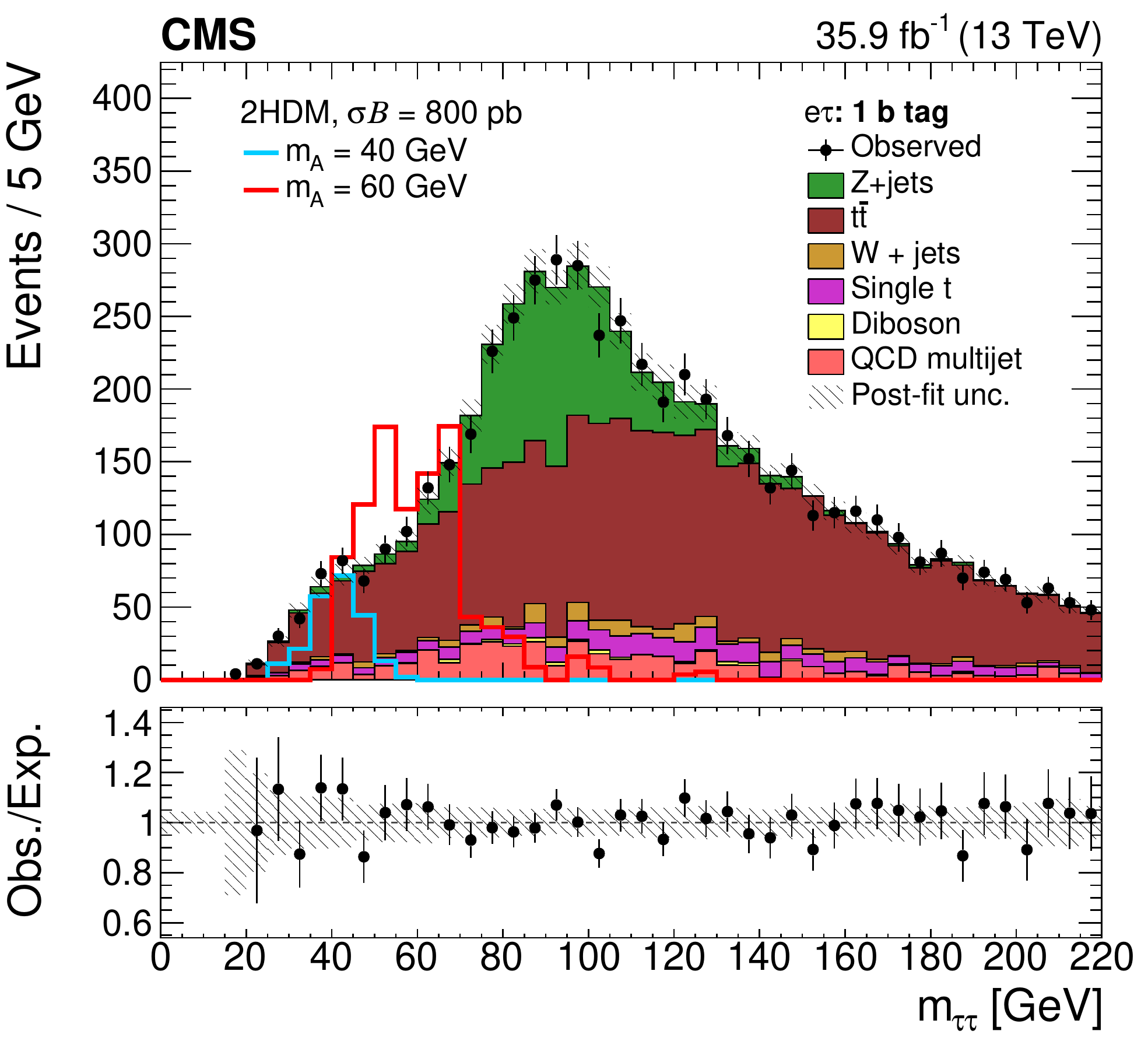}
  \includegraphics[width=\cmsFigWidth]{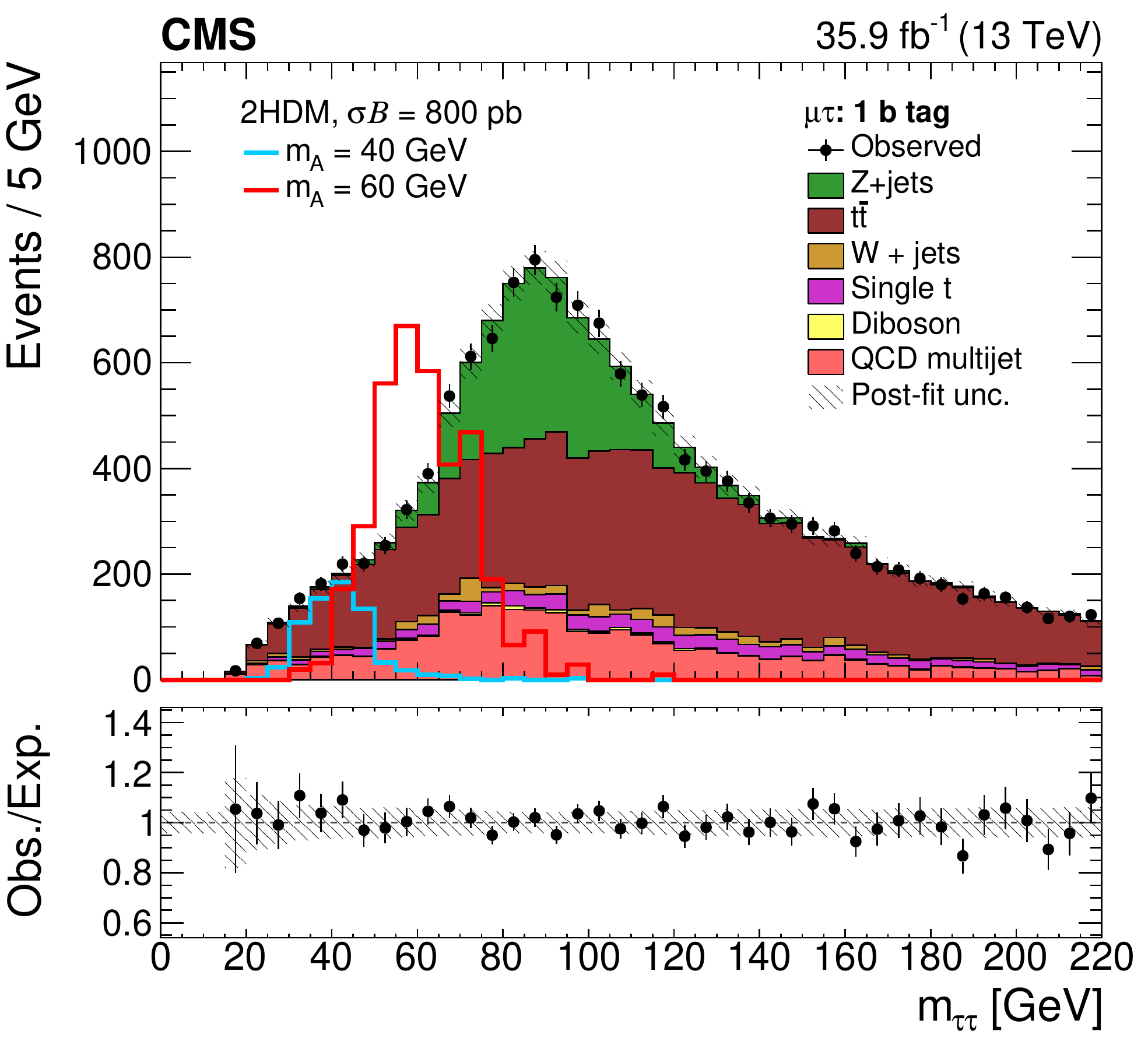}
  \caption{Measured $\mtt$ distribution in the $\etau$ (\cmsLeft), and $\mutau$ (\cmsRight) channel, compared to the expected SM background contributions. The signal distributions for $\bbA$ with a pseudoscalar mass of 40 and 60\GeV are overlaid to illustrate the sensitivity. They are normalized to the cross section times branching fraction of 800\unit{pb}. The uncertainty bands represent the sum in quadrature of statistical and systematic uncertainties obtained from the fit. The lower panels show the ratio between the observed and expected events in each bin.}
  \label{fig:5}
\end{figure}

\section{Results} \label{sec:results}

\begin{figure}[t!]
  \centering
  \includegraphics[width=\cmsFigWidth]{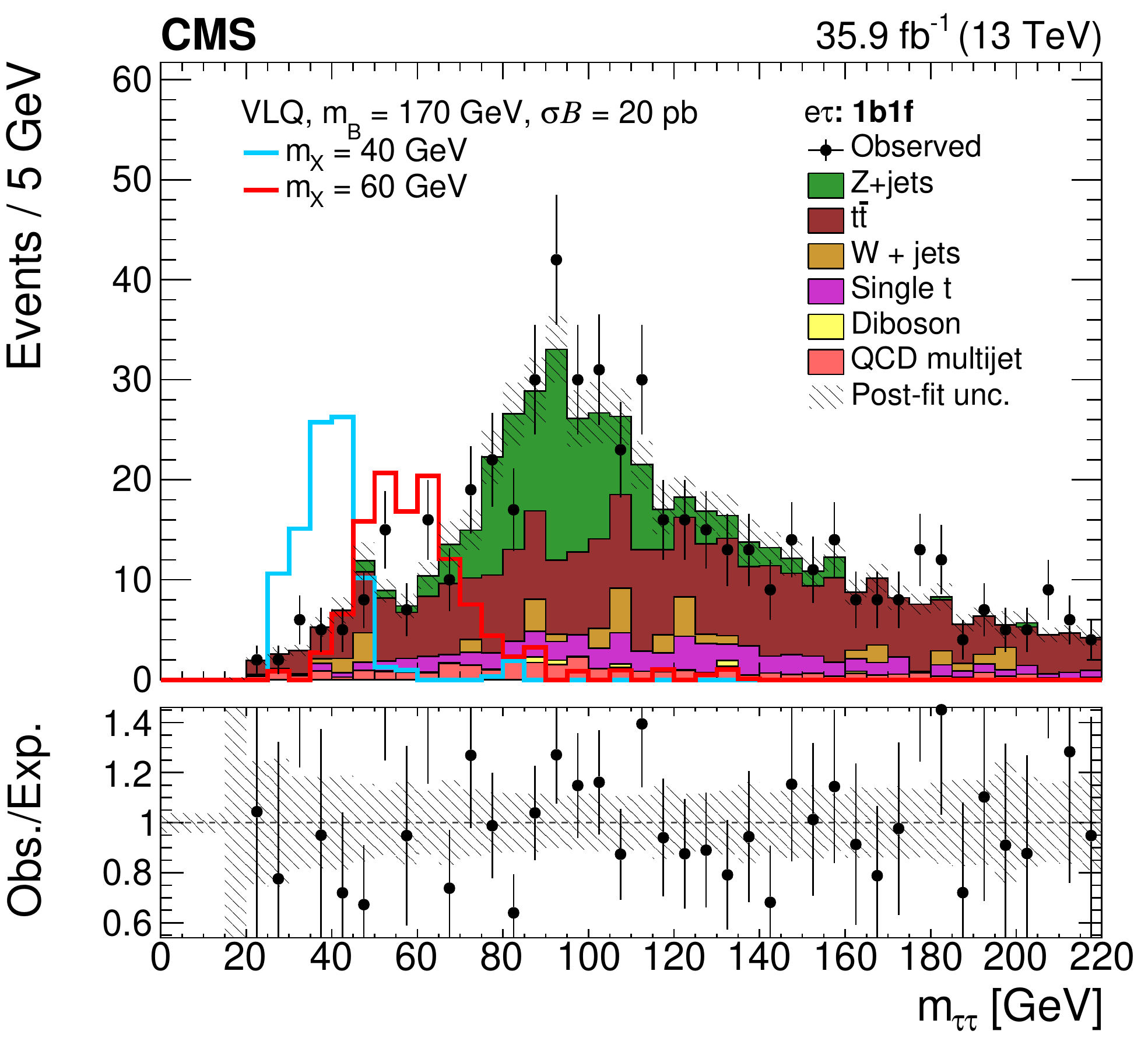}
  \includegraphics[width=\cmsFigWidth]{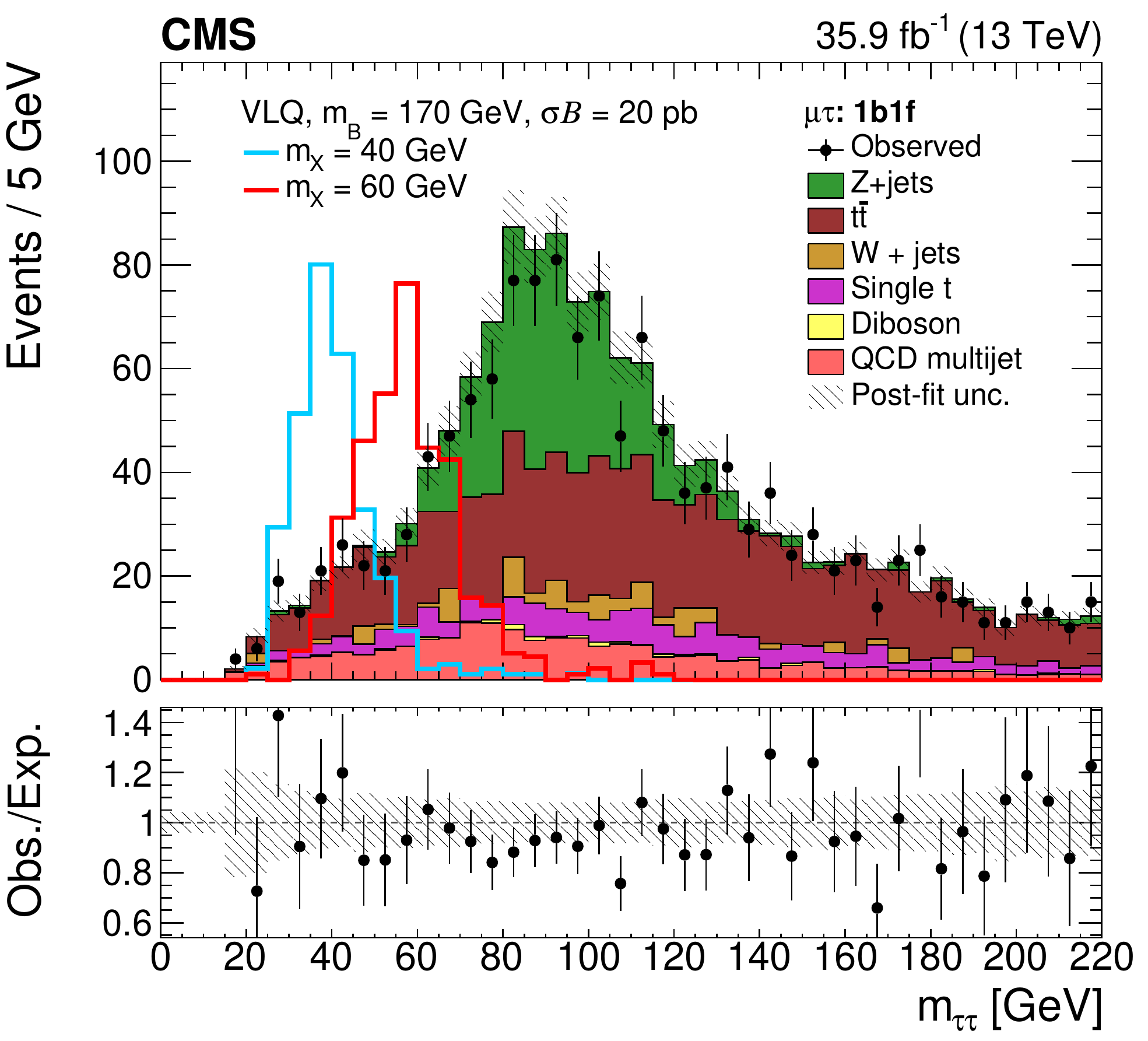}
  \includegraphics[width=\cmsFigWidth]{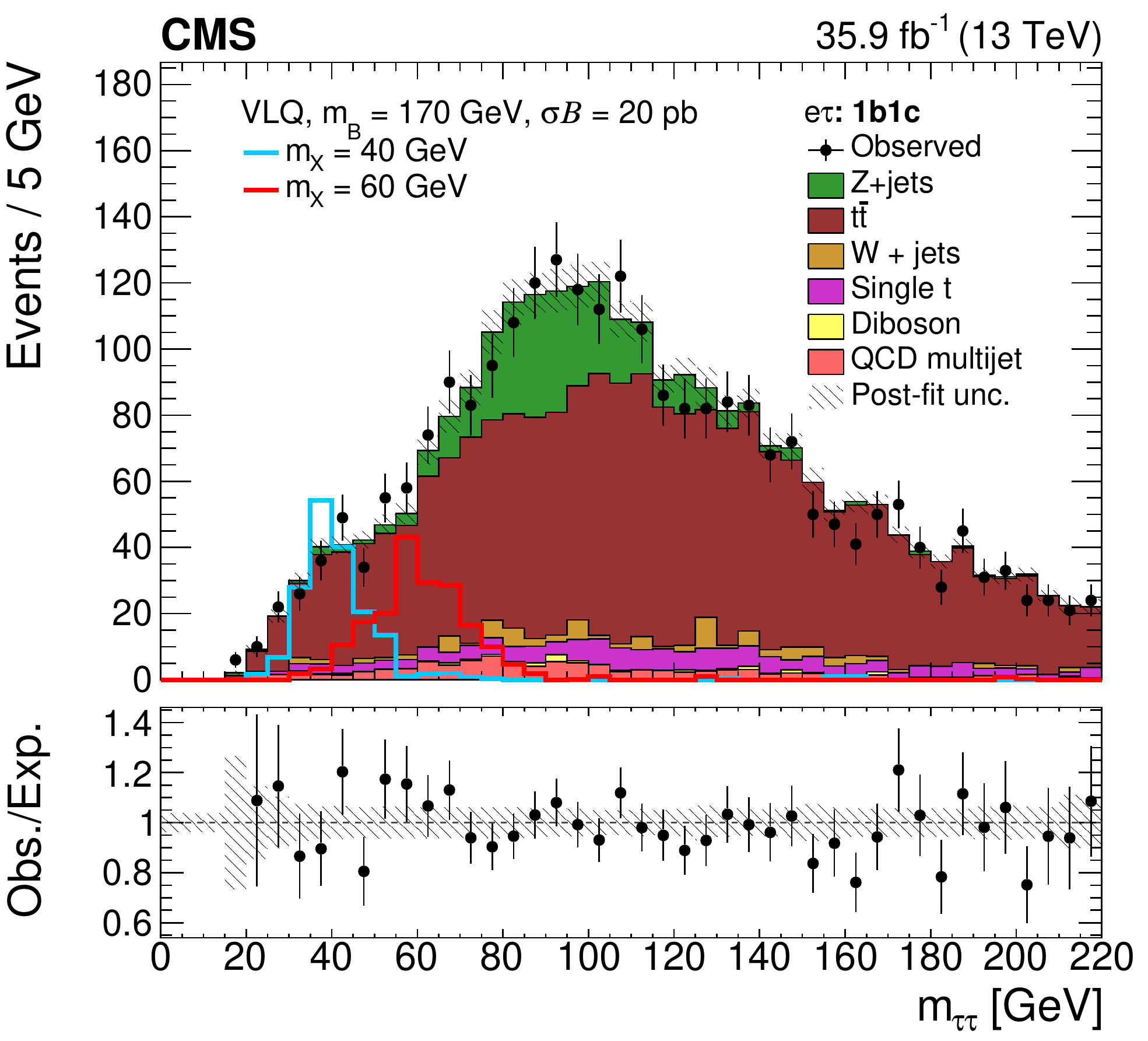}
  \includegraphics[width=\cmsFigWidth]{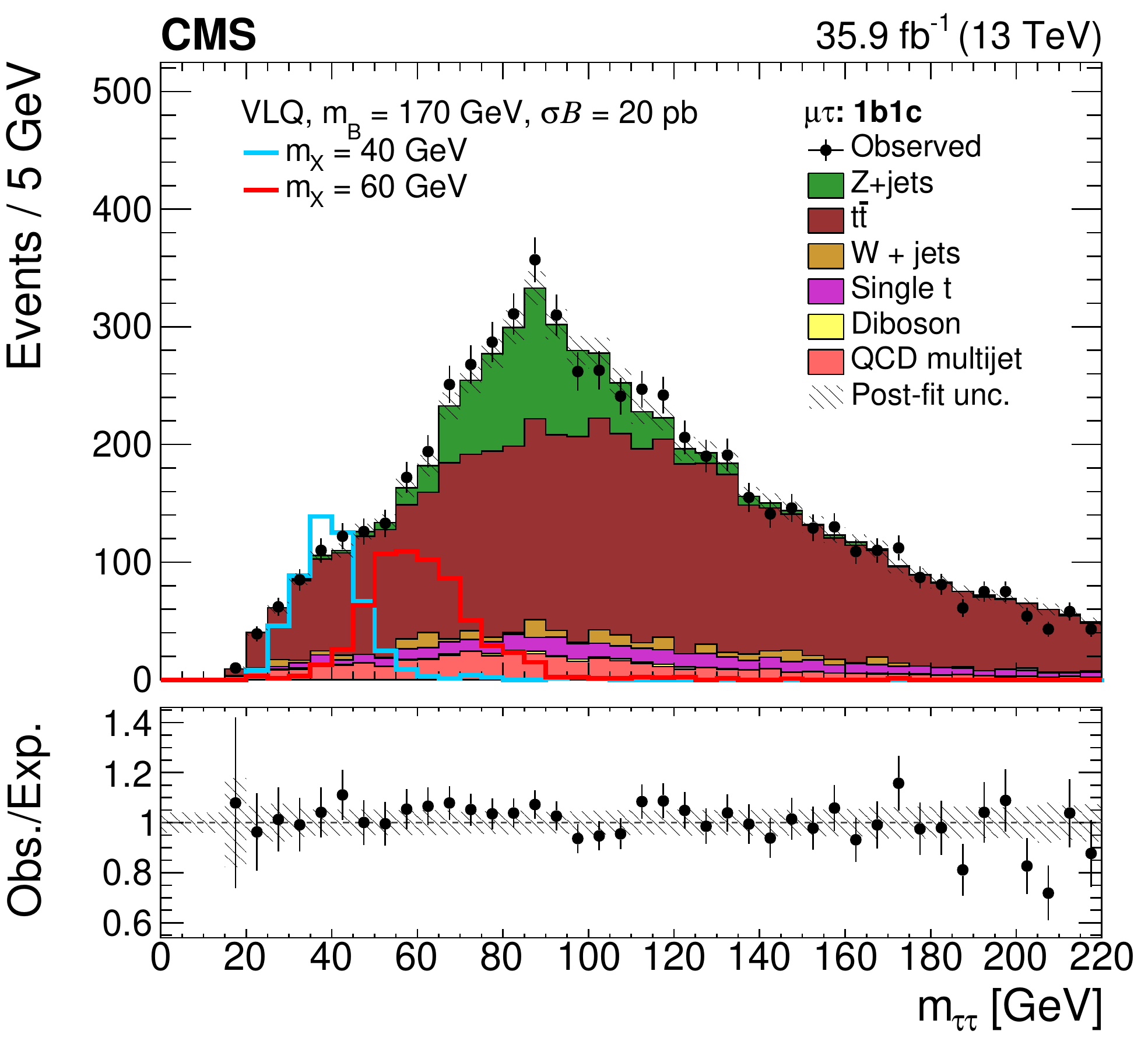}
  \caption{Measured $\mtt$ distribution in the $\etau$ (\cmsLeft), and $\mutau$ (\cmsRight) final states, for the 1b1f (upper) and 1b1c (lower) categories, compared to the expected SM background contributions. The signal distributions for the VLQ model with $\X$ boson masses of 40 and 60\GeV are overlaid to illustrate the sensitivity. They are normalized to the cross section times branching fraction of 20\unit{pb}. The uncertainty bands represent the sum in quadrature of statistical and systematic uncertainties obtained from the fit. The lower panels show the ratio between the observed and expected events in each bin.}
  \label{fig:6}
\end{figure}

Figure~\ref{fig:5} (\ref{fig:6}) shows the \SVFit mass distributions in the $\etau$ and $\mutau$ channel for the $\bbA$ (\qbX) search. Two signal contributions from a pseudoscalar (an \X boson) are overlaid assuming a mass of 40 or 60\GeV, normalized to an arbitrary cross section times branching fraction.
The uncertainty bands on the histograms of simulated events represent the sum in quadrature of statistical and systematic uncertainties, taking the full covariance matrix of all nuisance parameters into account.
However, uncertainties related to simulated events play a small role as the measurement is ultimately limited by the size of the data sample.

\begin{figure}[p]
    \centering
    \includegraphics[height=0.36\textwidth]{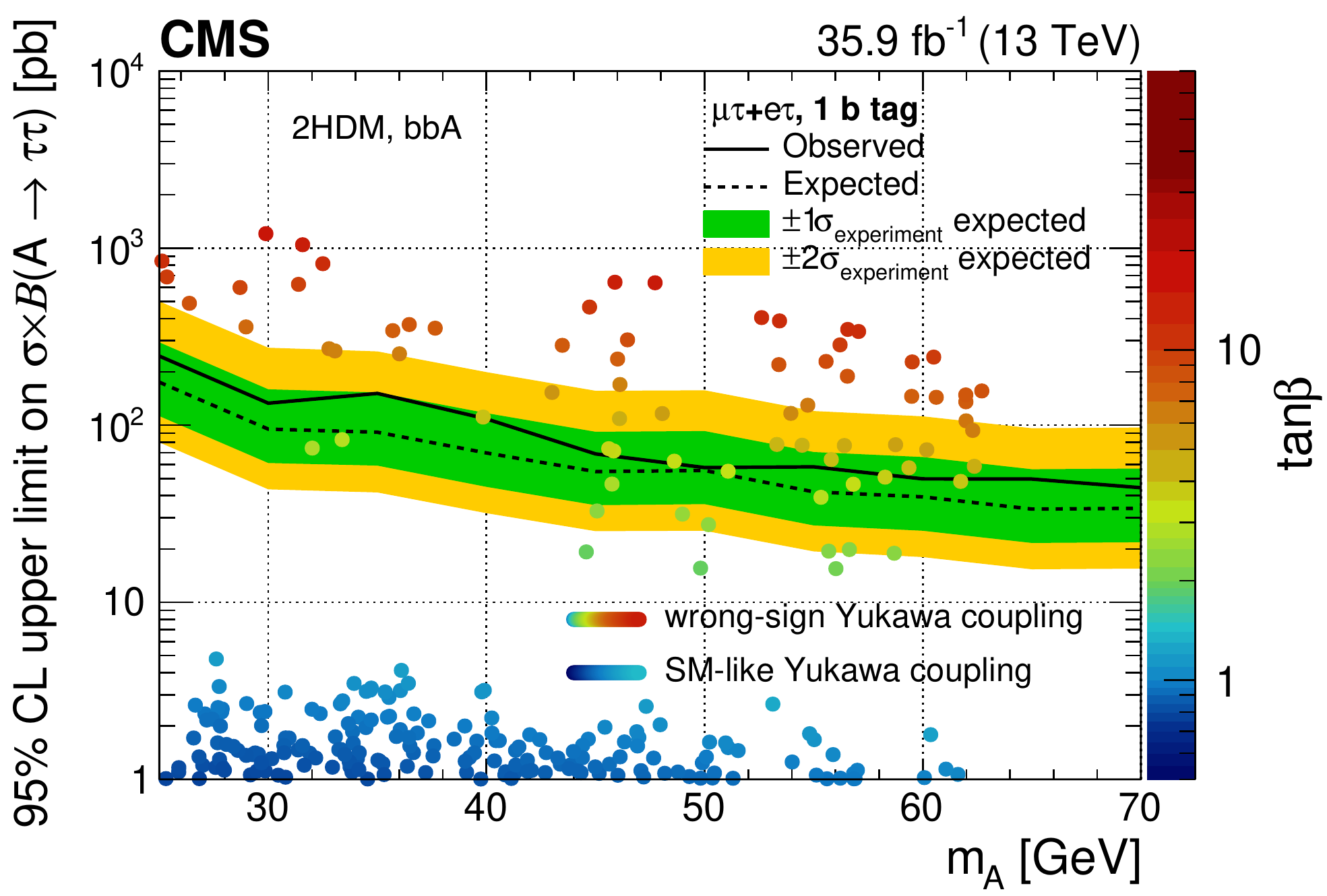}
    \caption{Observed (solid) and expected (dashed) limits at 95\% confidence level on the product of cross section for the production of the \bbA signal and branching fraction $\A\to\tautau$, obtained from the combination of the $\etau$ and $\mutau$ channels. The green and yellow bands represent the one and two standard deviation uncertainties in the expected limits. Representative 2HDMs with varied sets of the $\tan\beta$ and $\mA$ parameters are overlaid for two types of Yukawa couplings to the down-type fermions: one which is SM-like, and one in which the Yukawa coupling is negative (``wrong-sign'').}
\label{fig:limits_bbA}
\end{figure}

\begin{figure}[p]
    \centering
    \includegraphics[width=\cmsFigWidth]{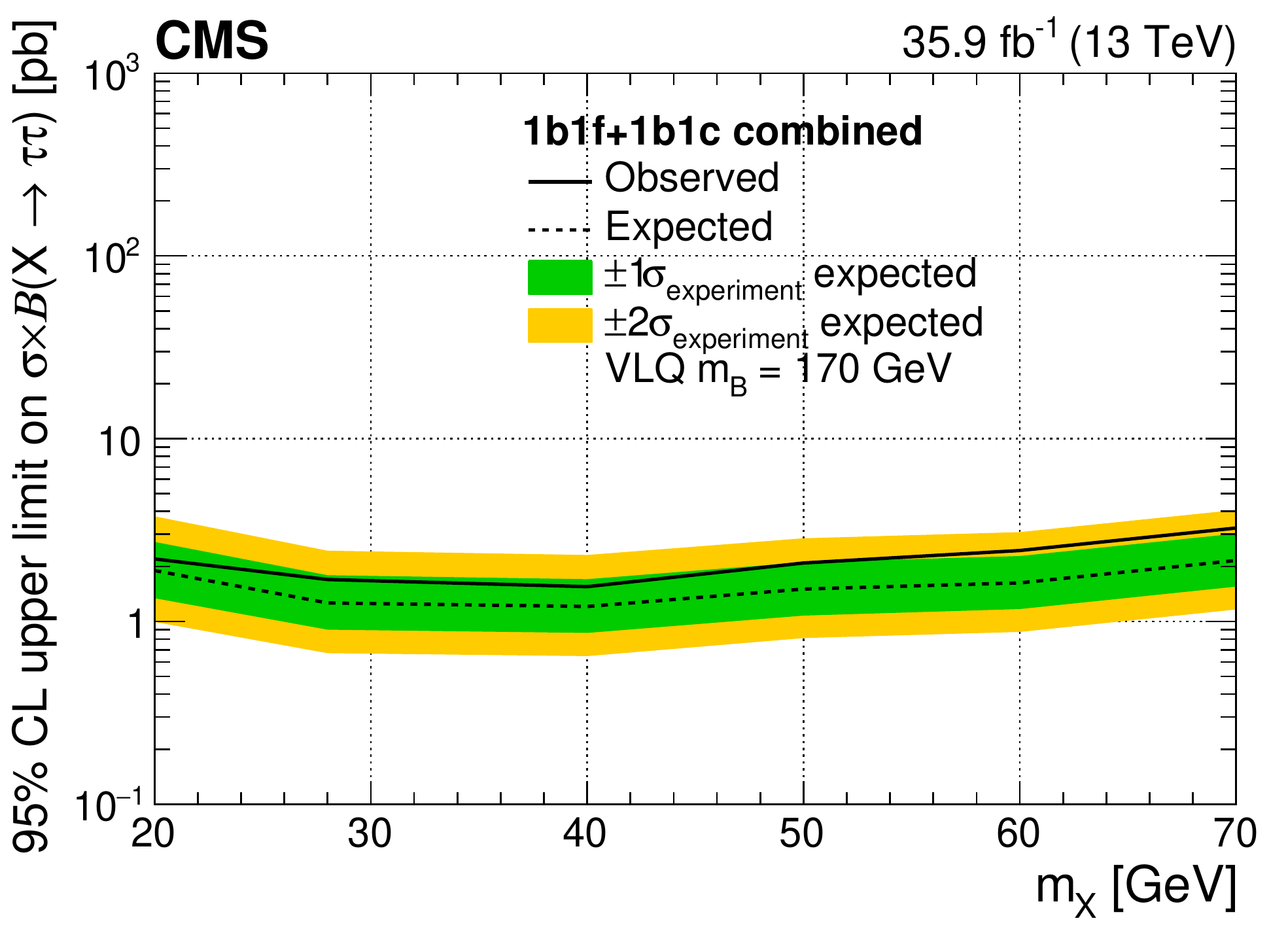}
    \includegraphics[width=\cmsFigWidth]{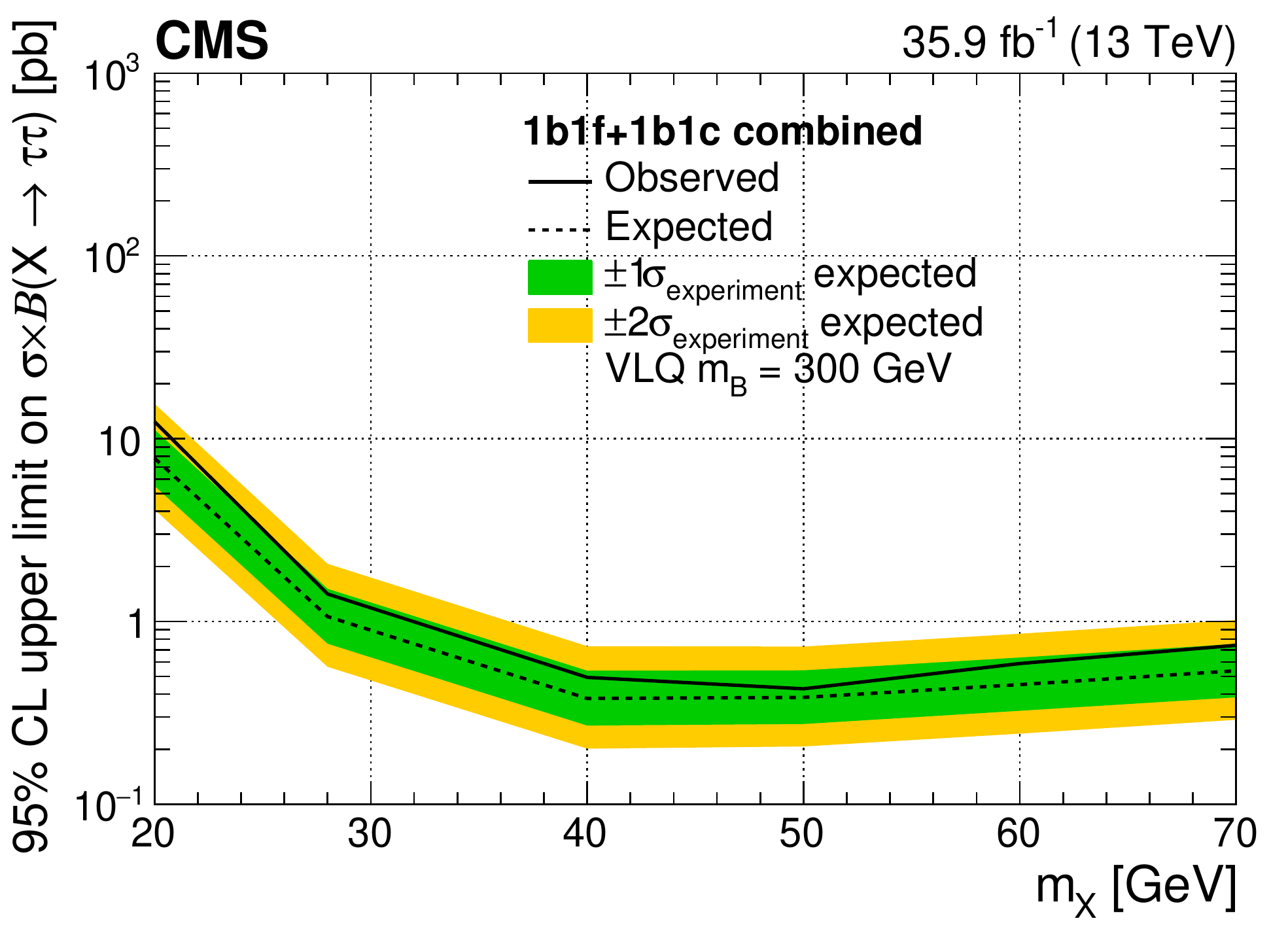}
    \includegraphics[width=\cmsFigWidth]{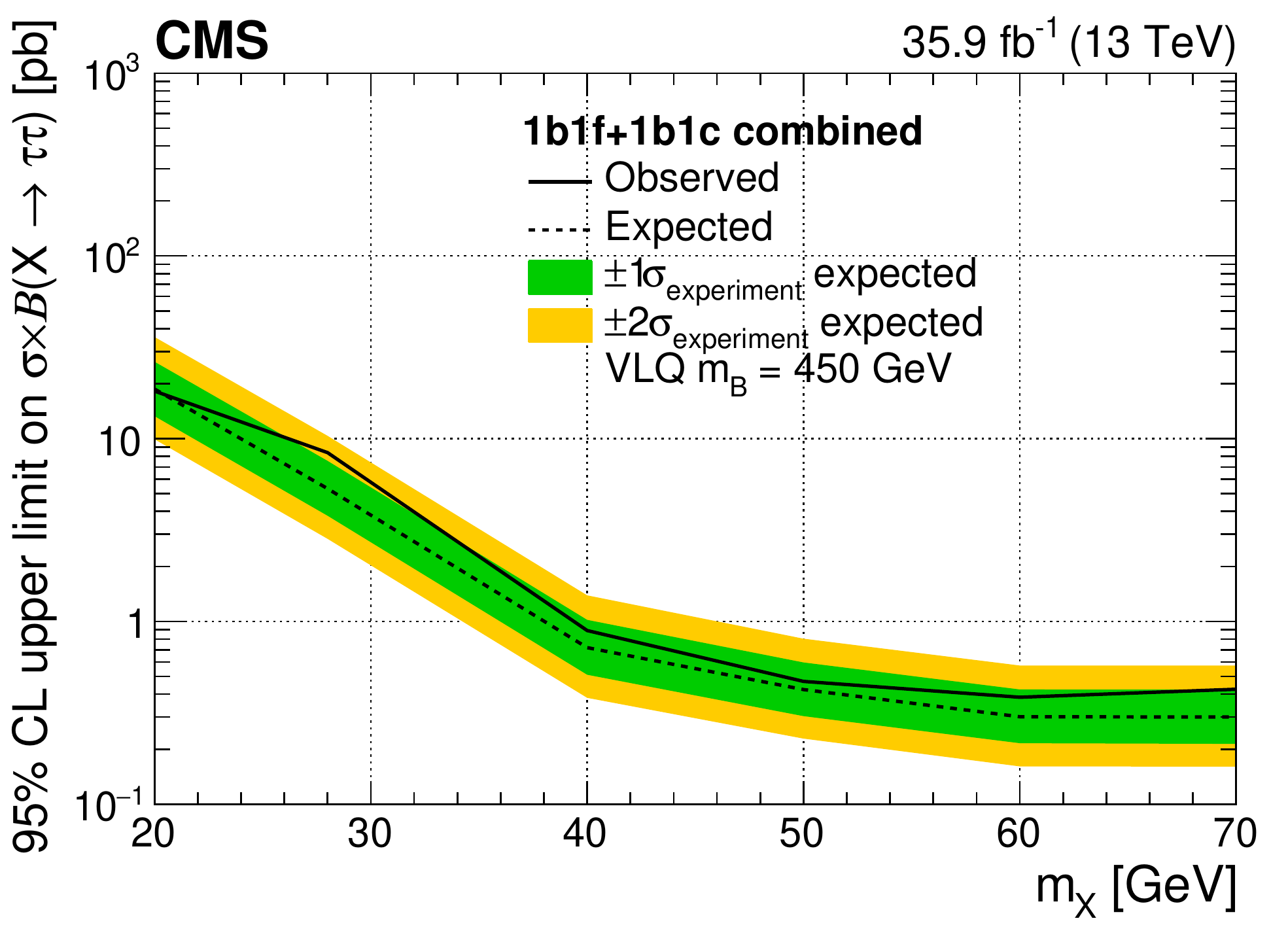}
    \caption{Observed (solid) and expected (dotted) limits at 95\% confidence level on the product of cross section for the production of the \qbX signal and branching fraction $\X\to\tautau$, obtained from the combination of the $\etau$ and $\mutau$ channels. The $\mB$ values of 170 (upper left), 300 (upper right), and 450\GeV are considered. The green and yellow bands represent the one and two standard deviation uncertainties in the expected limits.}
\label{fig:limits_vlq}
\end{figure}

The data are consistent with the background-only hypothesis of the SM, therefore, we set an upper limit on the cross section by using the asymptotic $\CLs$ modified-frequentist criterion~\cite{CLs1,Junk:1999kv,CLs3,LHC-HCG-Report}.
Figure~\ref{fig:limits_bbA} shows the observed and expected upper limits, at 95\% confidence level, on the cross section of \bbA production times branching fraction of $\A\to\tautau$ as a function of the pseudoscalar mass, $\mA$. Representative 2HDMs with varied sets of the $\tan\beta$ and $\mA$ parameters are also shown for two types of Yukawa couplings to the down-type fermions: one which is SM-like, and one in which the Yukawa coupling is negative and referred to as ``wrong-sign''~\cite{2HDM_8TeV}. We consider a $\tan\beta$ range of 0.6 to 2.0 (1.6 to 37) for the SM-like (wrong-sign) Yukawa coupling scenario with $\mA<65\GeV$. The cross sections for the wrong-sign Yukawa couplings are up to several orders of magnitude larger and have larger $\tan\beta$. Most of the cross sections for these models with $\tan\beta>3$ are excluded by the current data.
For signal events with an $\mA$ ranging from 30 to 70\GeV and \A decaying to a pair of {\Pgt} leptons, the efficiency to pass the final selection criteria of the 1 {\cPqb} tag category of the \mutau final state, including detector acceptance, selection efficiency, and branching fraction of $\A\to\tautau$, ranges from 0.002 to 0.022\%.
Figure~\ref{fig:limits_vlq} shows the same for the \qbX process in the VLQ model, but as a function of the \X boson mass $\mX$, for \B masses of 170, 300, and 450\GeV.
For both searches, the sensitivity is lower in the low-mass region because of the soft $\pt$ spectrum of the $\tauh$ candidate yielding a lower signal detection efficiency. In addition, as the boson mass decreases, the trajectories of the two {\Pgt} leptons are in close vicinity and start to spoil each other's isolation requirement. For the \qbX search, the 1b1f category drives the sensitivity, as can be inferred from Fig.~\ref{fig:6}.
For signal events in which $\mB=170\GeV$, with an \X mass ranging from 30 to 70\GeV and decaying to a pair of {\Pgt} leptons, the efficiency to pass the final selection criteria of the 1b1f category of the \mutau final state ranges from 0.03 to 0.06\%. These values range from 0.02 to 0.10\% for the same final state of the 1b1c category.

We proceed to make a comparison with Ref.~\cite{dimuon_CMS}, that is based on the same data set as this paper, and defines two similar signal event categories, but with a dimuon pair in the final state instead of a {\Pgt}-lepton pair. Upper limits are set at 95\% confidence level on the fiducial cross section for the production of a 28\GeV particle decaying to two muons. Because the analysis does not consider a signal model that specifies the kinematic acceptance, it defines the fiducial cross section as
\begin{equation*}\label{eq:fid}
  \sigma_\text{fid} = \frac{\NS}{L\epsmumu},
\end{equation*}
where $\NS$ is the number of signal events extracted from the fit to the dimuon mass spectrum, $L$ is the integrated luminosity, and $\epsmumu=0.28$ is the reconstruction efficiency, which takes into account the muon trigger, identification and isolation, as well as the {\cPqb}-tagging efficiency.
To compare these results to the present analysis with a {\Pgt}-lepton pair in the final state, we consider only the most sensitive final state, $\mutau$. The reconstruction efficiency $\epsmutau$ for this final state is estimated to be $0.10$. This includes the muon trigger, identification and isolation, as well as the $\tauh$ identification and {\cPqb} tagging efficiency.
Taking into account $\epsmutau$, the upper limit on the fiducial cross section is 0.029 (0.057)\unit{pb} for 1b1f (1b1c), while for the dimuon search, the upper limit is 0.0037 (0.0032)\unit{pb} for similar event categories.
As expected, this analysis is less sensitive than the dimuon search to a hypothetical signal that decays equally to all flavors of leptons. However, if there were a Yukawa-type enhancement between the signal and the {\Pgt} leptons, then the constraints on the signal production cross section by this analysis would improve by a factor of $m_{\Pgt}^2/m_{\Pgm}^2$.

\section{Summary} \label{sec:conclusions}

This paper presents a general search for a low-mass $\TT$ resonance produced in association with a bottom quark.
After defining the signal region by the presence of an electron or muon consistent with the decay of a {\Pgt} lepton, a hadronically decaying {\Pgt} lepton, and a jet originating from a bottom quark, an excess over standard model background is searched for in the reconstructed invariant mass distribution of the inferred \tautau system.
The data are consistent with the standard model background.
We set upper limits at 95\% confidence level on the cross section times branching fraction for two signal models: a light pseudoscalar Higgs boson decaying to a pair of {\Pgt} leptons produced in association with a bottom quark, and a low-mass boson \X decaying to a {\Pgt}-lepton pair that is produced in the decay of a bottom-like quark $\B$ as $\B\to\cPqb\X$. For both scenarios, \X boson masses between 20 and 70\GeV are probed. Upper limits at 95\% confidence level ranging from 250 to 44\unit{pb} are set on the light pseudoscalar, and from 20 to 0.3\unit{pb} on $\B$ masses between 170 and 450\GeV.
This is the first search for an \X resonance in this final state using the center-of-mass energy of 13\TeV.
Since many extensions of the standard model have similar event kinematics as this analysis, these results could also be applied to put constraints on other low-mass \tautau resonances. If there were a Yukawa-type enhancement between the signal and the {\Pgt} leptons, then the constraints on the signal production cross section by this analysis would improve by a factor of $m_{\Pgt}^2/m_{\Pgm}^2$.

The optimized selection of this analysis targets previously unexplored decays of heavy bottom-like quarks, providing new sensitivity to vector-like quarks.

\begin{acknowledgments}
We congratulate our colleagues in the CERN accelerator departments for the excellent performance of the LHC and thank the technical and administrative staffs at CERN and at other CMS institutes for their contributions to the success of the CMS effort. In addition, we gratefully acknowledge the computing centres and personnel of the Worldwide LHC Computing Grid for delivering so effectively the computing infrastructure essential to our analyses. Finally, we acknowledge the enduring support for the construction and operation of the LHC and the CMS detector provided by the following funding agencies: BMBWF and FWF (Austria); FNRS and FWO (Belgium); CNPq, CAPES, FAPERJ, FAPERGS, and FAPESP (Brazil); MES (Bulgaria); CERN; CAS, MoST, and NSFC (China); COLCIENCIAS (Colombia); MSES and CSF (Croatia); RPF (Cyprus); SENESCYT (Ecuador); MoER, ERC IUT, PUT and ERDF (Estonia); Academy of Finland, MEC, and HIP (Finland); CEA and CNRS/IN2P3 (France); BMBF, DFG, and HGF (Germany); GSRT (Greece); NKFIA (Hungary); DAE and DST (India); IPM (Iran); SFI (Ireland); INFN (Italy); MSIP and NRF (Republic of Korea); MES (Latvia); LAS (Lithuania); MOE and UM (Malaysia); BUAP, CINVESTAV, CONACYT, LNS, SEP, and UASLP-FAI (Mexico); MOS (Montenegro); MBIE (New Zealand); PAEC (Pakistan); MSHE and NSC (Poland); FCT (Portugal); JINR (Dubna); MON, RosAtom, RAS, RFBR, and NRC KI (Russia); MESTD (Serbia); SEIDI, CPAN, PCTI, and FEDER (Spain); MOSTR (Sri Lanka); Swiss Funding Agencies (Switzerland); MST (Taipei); ThEPCenter, IPST, STAR, and NSTDA (Thailand); TUBITAK and TAEK (Turkey); NASU and SFFR (Ukraine); STFC (United Kingdom); DOE and NSF (USA).

\hyphenation{Rachada-pisek} Individuals have received support from the Marie-Curie program and the European Research Council and Horizon 2020 Grant, contract Nos.\ 675440 and 765710 (European Union); the Leventis Foundation; the A.P.\ Sloan Foundation; the Alexander von Humboldt Foundation; the Belgian Federal Science Policy Office; the Fonds pour la Formation \`a la Recherche dans l'Industrie et dans l'Agriculture (FRIA-Belgium); the Agentschap voor Innovatie door Wetenschap en Technologie (IWT-Belgium); the F.R.S.-FNRS and FWO (Belgium) under the ``Excellence of Science -- EOS" -- be.h project n.\ 30820817; the Beijing Municipal Science \& Technology Commission, No. Z181100004218003; the Ministry of Education, Youth and Sports (MEYS) of the Czech Republic; the Lend\"ulet (``Momentum") Program and the J\'anos Bolyai Research Scholarship of the Hungarian Academy of Sciences, the New National Excellence Program \'UNKP, the NKFIA research grants 123842, 123959, 124845, 124850, 125105, 128713, 128786, and 129058 (Hungary); the Council of Science and Industrial Research, India; the HOMING PLUS program of the Foundation for Polish Science, cofinanced from European Union, Regional Development Fund, the Mobility Plus program of the Ministry of Science and Higher Education, the National Science Center (Poland), contracts Harmonia 2014/14/M/ST2/00428, Opus 2014/13/B/ST2/02543, 2014/15/B/ST2/03998, and 2015/19/B/ST2/02861, Sonata-bis 2012/07/E/ST2/01406; the National Priorities Research Program by Qatar National Research Fund; the Programa Estatal de Fomento de la Investigaci{\'o}n Cient{\'i}fica y T{\'e}cnica de Excelencia Mar\'{\i}a de Maeztu, grant MDM-2015-0509 and the Programa Severo Ochoa del Principado de Asturias; the Thalis and Aristeia programs cofinanced by EU-ESF and the Greek NSRF; the Rachadapisek Sompot Fund for Postdoctoral Fellowship, Chulalongkorn University and the Chulalongkorn Academic into Its 2nd Century Project Advancement Project (Thailand); the Welch Foundation, contract C-1845; and the Weston Havens Foundation (USA).
\end{acknowledgments}
\bibliography{auto_generated}

\ifthenelse{\boolean{cms@external}}{}{
\clearpage
\appendix
\numberwithin{table}{section}
\numberwithin{figure}{section}
\section{Supplementary material}\label{app:suppMat}

\input{supplemental_material}

}

\cleardoublepage \section{The CMS Collaboration \label{app:collab}}\begin{sloppypar}\hyphenpenalty=5000\widowpenalty=500\clubpenalty=5000\input{HIG-17-014-authorlist.tex}\end{sloppypar}
\end{document}

%% file: supplemental_material.tex
\begin{figure}[h]
  \centering
  \includegraphics[width=\cmsFigWidth]{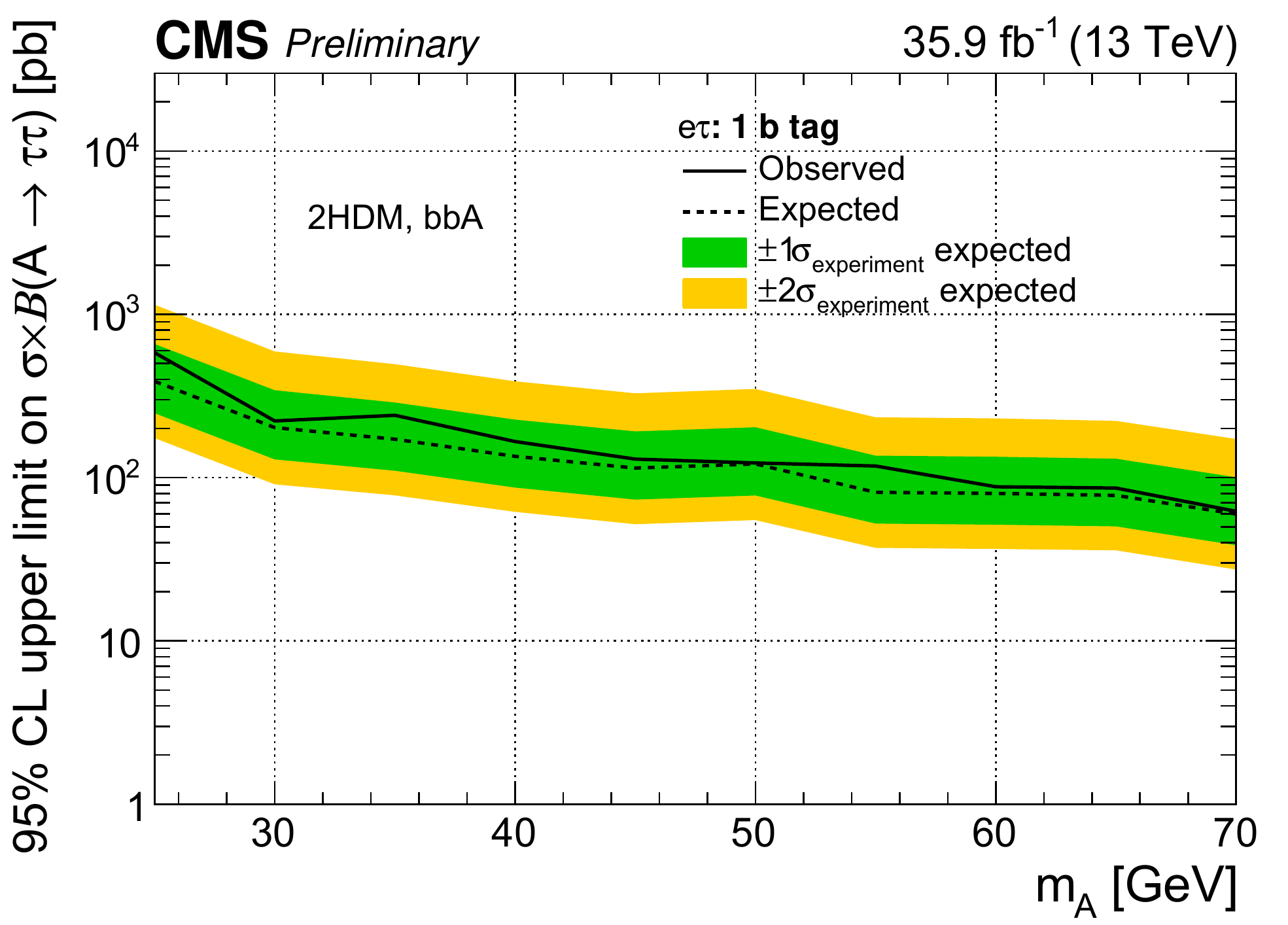}
  \includegraphics[width=\cmsFigWidth]{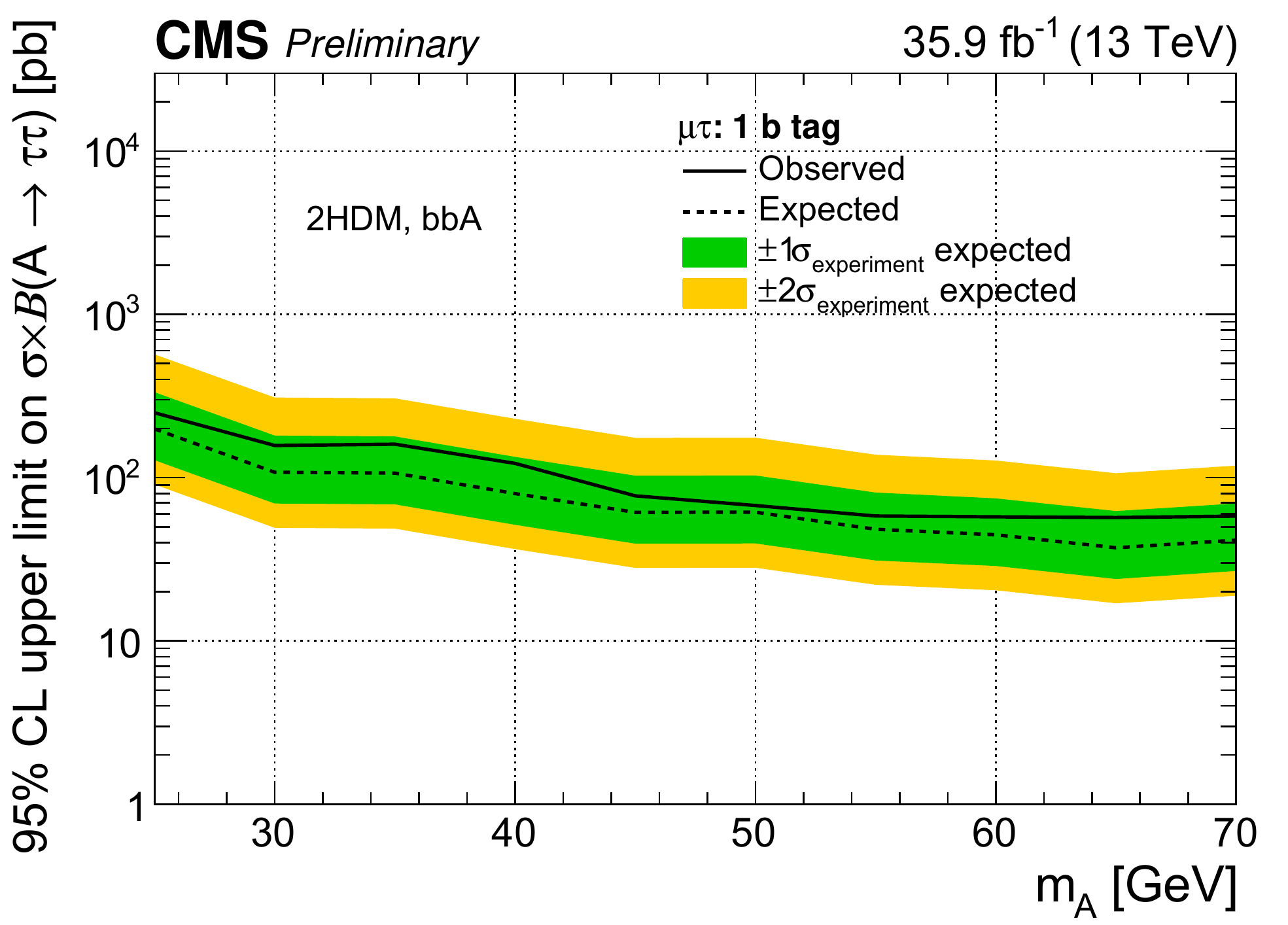}
  \caption{Observed (solid) and expected (dashed) limits at 95\% confidence level on the product of cross section for the production of the \bbA signal and branching fraction $\A\to\tautau$, obtained for the $\etau$ (left) and $\mutau$ (right) channels. The green and yellow bands represent the one and two standard deviation uncertainties in the expected limits.}
\end{figure}

\begin{table}[h]
\centering
\topcaption{The product of acceptance, efficiency, and branching fraction of the $\bbA$ signal with $\A\to\tautau$ in the $\mutau$ and $\etau$ channels of the 1 b tag category, for different \A boson mass values. The selections are as described in Section~\ref{sec:selection}. The uncertainty refers to the statistical uncertainty only.}
\begin{tabular}{ccc}
    \hline
    $\mA$ [\GeVns{}] & $\etau$ [\%] & $\mutau$ [\%] \\
  \hline
    25  &  $0.00025 \pm 0.00002$  &  $0.00097 \pm 0.00005$  \\
    30  &  $0.00066 \pm 0.00004$  &  $0.00226 \pm 0.00008$  \\
    40  &  $0.00160 \pm 0.00006$  &  $0.0043  \pm 0.0001$  \\
    50  &  $0.00229 \pm 0.00008$  &  $0.0078  \pm 0.0001$  \\
    60  &  $0.0046  \pm 0.0001$   &  $0.0144  \pm 0.0002$  \\
    70  &  $0.0087  \pm 0.0001$   &  $0.0222  \pm 0.0002$  \\
  \hline
\end{tabular}
\end{table}

\begin{figure}[p]
  \centering
  \includegraphics[width=\cmsFigWidth]{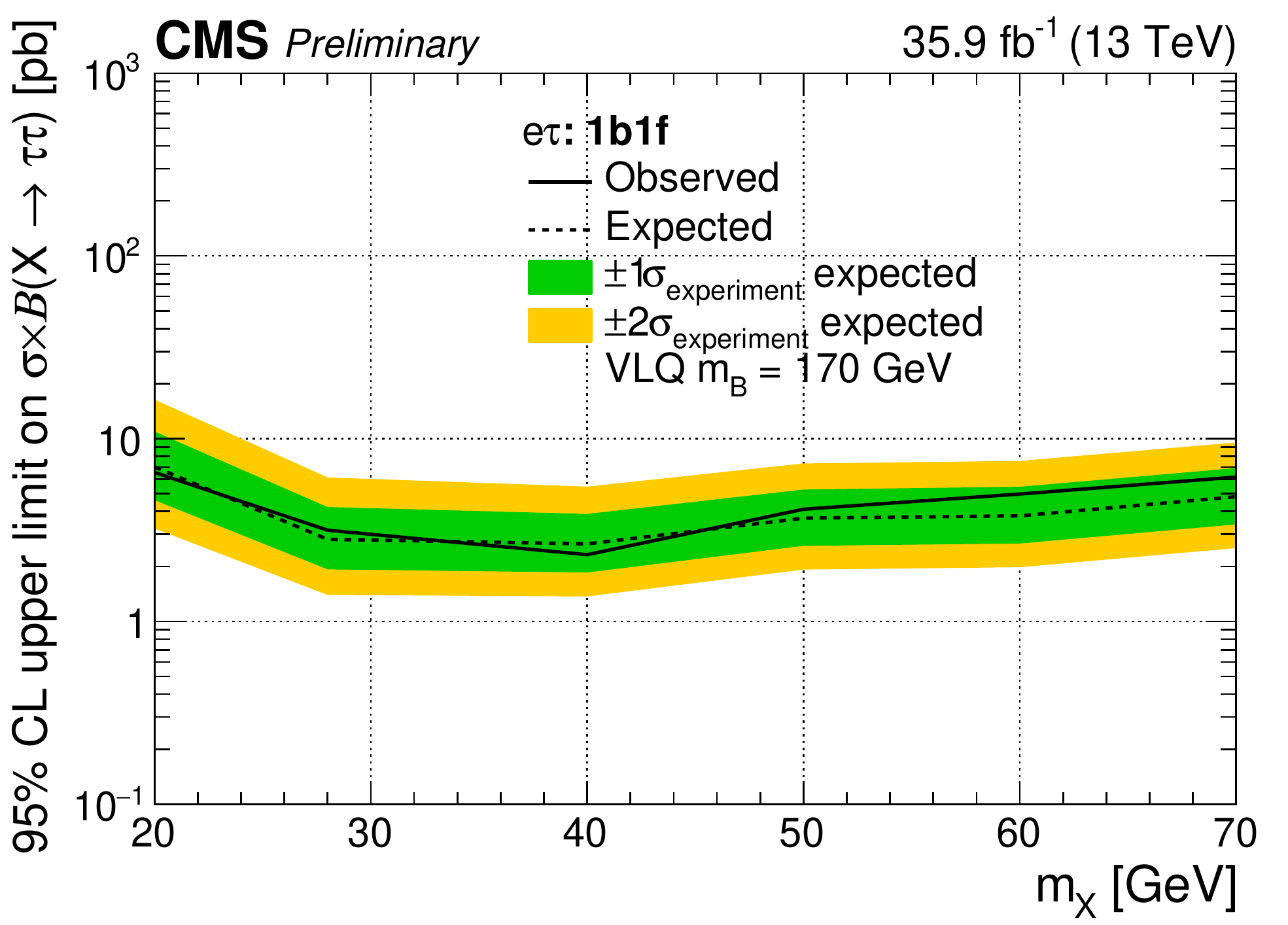}
  \includegraphics[width=\cmsFigWidth]{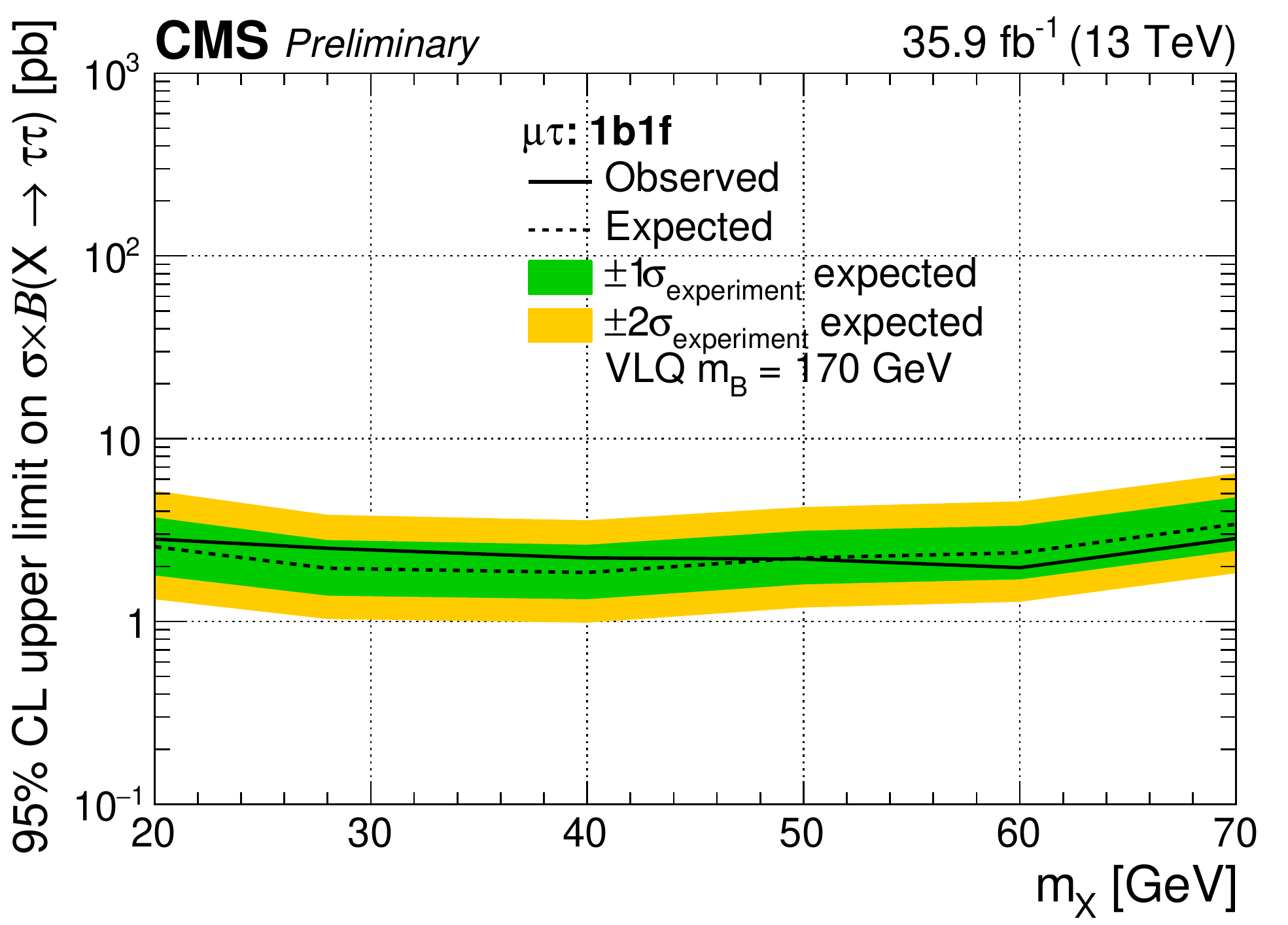}
  \includegraphics[width=\cmsFigWidth]{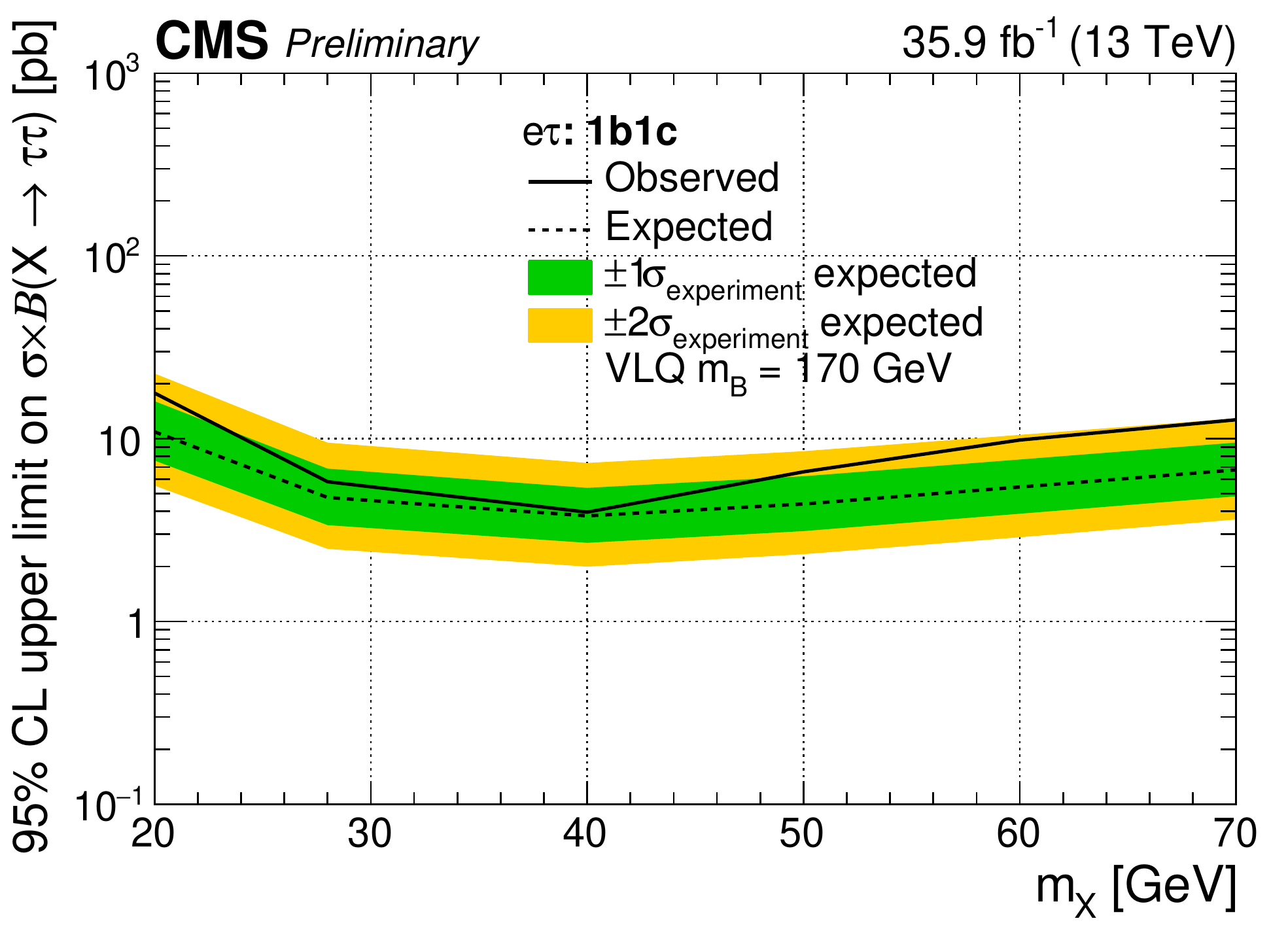}
  \includegraphics[width=\cmsFigWidth]{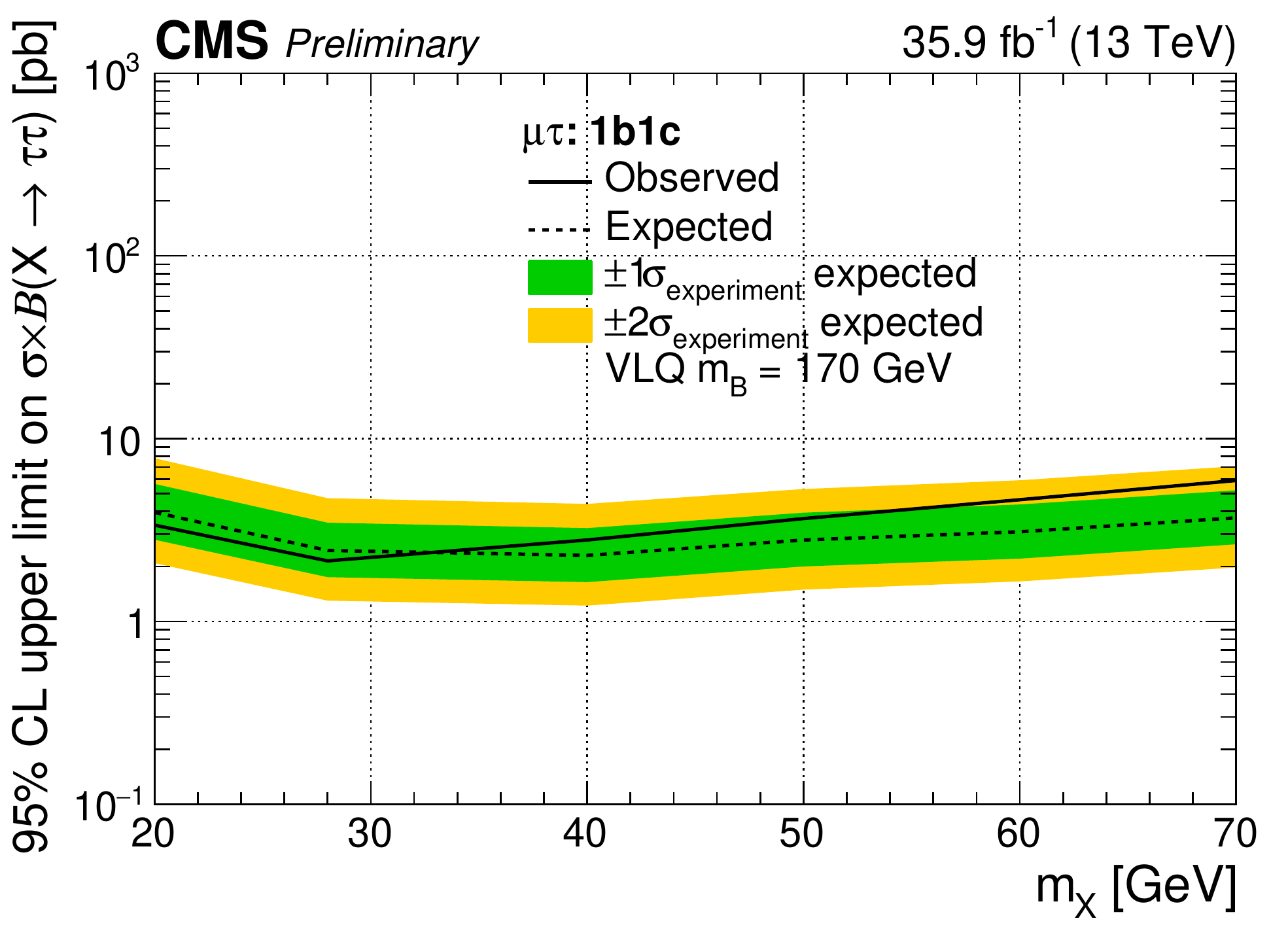}
  \caption{Observed (solid) and expected (dotted) limits at 95\% confidence level on the product of cross section for the production of the \qbX signal and branching fraction $\X\to\tautau$, obtained for the $\etau$ (left) and $\mutau$ (right) channels in the 1b1f (top) and 1b1c (bottom) categories. The scenario with $\mB=170\GeV$ is considered. The green and yellow bands represent the one and two standard deviation uncertainties in the expected limits.}
\end{figure}

\begin{table}[p]
\centering
\topcaption{The product of acceptance, efficiency, and branching fraction of the $\qbX$ signal with $\X\to\tautau$ in the $\mutau$ and $\etau$ channels of the 1b1f and 1b1f categories, for different \X boson mass values. The selections are as described in Section~\ref{sec:selection}. The uncertainty refers to the statistical uncertainty only.}
\begin{tabular}{@{\extracolsep{4pt}}ccccc@{}}
  \hline
  \multirow{2}{*}{$\mX$ [\GeVns{}]} & \multicolumn{2}{c}{1b1f} & \multicolumn{2}{c}{1b1c} \\
                                 \cline{2-3}                \cline{4-5}
        & $\etau$ [\%] & $\mutau$ [\%] & $\etau$ [\%] & $\mutau$ [\%] \\
  \hline
    20  &  $0.0037 \pm 0.0001$  &  $0.0146 \pm 0.0001$  &  $0.0044 \pm 0.0001$  &  $0.0181 \pm 0.0001$  \\
    30  &  $0.0098 \pm 0.0001$  &  $0.0293 \pm 0.0001$  &  $0.0165 \pm 0.0001$  &  $0.0496 \pm 0.0002$  \\
    40  &  $0.0162 \pm 0.0002$  &  $0.0466 \pm 0.0003$  &  $0.0307 \pm 0.0002$  &  $0.0765 \pm 0.0004$  \\
    50  &  $0.0183 \pm 0.0002$  &  $0.0494 \pm 0.0003$  &  $0.0321 \pm 0.0002$  &  $0.0844 \pm 0.0004$  \\
    60  &  $0.0212 \pm 0.0002$  &  $0.0531 \pm 0.0003$  &  $0.0331 \pm 0.0003$  &  $0.0957 \pm 0.0004$  \\
    70  &  $0.0225 \pm 0.0003$  &  $0.0562 \pm 0.0004$  &  $0.0375 \pm 0.0003$  &  $0.0991 \pm 0.0005$  \\
\hline
\end{tabular}
\end{table}

%% file: HIG-17-014-authorlist.tex
\vskip\cmsinstskip
\textbf{Yerevan Physics Institute, Yerevan, Armenia}\\*[0pt]
A.M.~Sirunyan, A.~Tumasyan
\vskip\cmsinstskip
\textbf{Institut f\"{u}r Hochenergiephysik, Wien, Austria}\\*[0pt]
W.~Adam, F.~Ambrogi, E.~Asilar, T.~Bergauer, J.~Brandstetter, M.~Dragicevic, J.~Er\"{o}, A.~Escalante~Del~Valle, M.~Flechl, R.~Fr\"{u}hwirth\cmsAuthorMark{1}, V.M.~Ghete, J.~Hrubec, M.~Jeitler\cmsAuthorMark{1}, N.~Krammer, I.~Kr\"{a}tschmer, D.~Liko, T.~Madlener, I.~Mikulec, N.~Rad, H.~Rohringer, J.~Schieck\cmsAuthorMark{1}, R.~Sch\"{o}fbeck, M.~Spanring, D.~Spitzbart, W.~Waltenberger, J.~Wittmann, C.-E.~Wulz\cmsAuthorMark{1}, M.~Zarucki
% ...
\vskip\cmsinstskip
% ...
1:  Also at Vienna University of Technology, Vienna, Austria\\
% ...


%% file: HIG-17-014_temp.bbl
\providecommand{\href}[2]{#2}\begingroup\raggedright\begin{thebibliography}{10}%
\makeatletter
\providecommand{\hrefCMSnoop }[0]{\@secondoftwo}%
\makeatother
\providecommand{\doi}{\texttt{doi:}\begingroup \urlstyle{tt}\Url}

\bibitem{Higgs_discovery_2012_ATLAS}
\hrefCMSnoop {}{{ATLAS Collaboration}, ``{Observation of a new particle in the
  search for the Standard Model Higgs boson with the ATLAS detector at the
  LHC}'',} \textit{ Phys. Lett. B} \textbf{ 716} (2012) 1,
  \href{http://dx.doi.org/10.1016/j.physletb.2012.08.020}{\doi{10.1016/j.physletb.2012.08.020}},
\href{http://www.arXiv.org/abs/1207.7214}{\texttt{arXiv:1207.7214}}.

\bibitem{Higgs_discovery_2012_CMS}
\hrefCMSnoop {}{{CMS Collaboration}, ``{Observation of a new boson at a mass of
  125 GeV with the CMS experiment at the LHC}'',} \textit{ Phys. Lett. B}
  \textbf{ 716} (2012) 30,
  \href{http://dx.doi.org/10.1016/j.physletb.2012.08.021}{\doi{10.1016/j.physletb.2012.08.021}},
\href{http://www.arXiv.org/abs/1207.7235}{\texttt{arXiv:1207.7235}}.

\bibitem{Higgs_discovery_2013_CMS}
\hrefCMSnoop {}{{CMS Collaboration}, ``{Observation of a new boson with mass
  near 125 {GeV} in pp collisions at $\sqrt{s}$ = 7 and 8 {TeV}}'',} \textit{
  JHEP} \textbf{ 06} (2013) 081,
  \href{http://dx.doi.org/10.1007/JHEP06(2013)081}{\doi{10.1007/JHEP06(2013)081}},
\href{http://www.arXiv.org/abs/1303.4571}{\texttt{arXiv:1303.4571}}.

\bibitem{SM1}
\hrefCMSnoop {}{S.~Weinberg, ``A model of leptons'',} \textit{ Phys. Rev.
  Lett.} \textbf{ 19} (1967) 1264,
\href{http://dx.doi.org/10.1103/PhysRevLett.19.1264}{\doi{10.1103/PhysRevLett.19.1264}}.

\bibitem{Higgs_theory1}
\hrefCMSnoop {}{F.~Englert and R.~Brout, ``Broken symmetry and the mass of
  gauge vector mesons'',} \textit{ Phys. Rev. Lett.} \textbf{ 13} (1964) 321,
\href{http://dx.doi.org/10.1103/PhysRevLett.13.321}{\doi{10.1103/PhysRevLett.13.321}}.

\bibitem{Higgs_theory2}
\hrefCMSnoop {}{P.~W. Higgs, ``Broken symmetries and the masses of gauge
  bosons'',} \textit{ Phys. Rev. Lett.} \textbf{ 13} (1964) 508,
  \href{http://dx.doi.org/10.1103/PhysRevLett.13.508}{\doi{10.1103/PhysRevLett.13.508}}.

\bibitem{ATLAS-MASS}
\hrefCMSnoop {}{{ATLAS Collaboration}, ``{{Measurement of the Higgs boson mass
  from the $\mathrm{H}\rightarrow \gamma\gamma$ and $\mathrm{H}\rightarrow
  \mathrm{Z}\mathrm{Z}^* \rightarrow 4\ell$ channels in pp collisions at
  center-of-mass energies 7 and 8\TeV with the ATLAS detector}}'',} \textit{
  Phys. Rev. D} \textbf{ 90} (2014) 052004,
  \href{http://dx.doi.org/10.1103/PhysRevD.90.052004}{\doi{10.1103/PhysRevD.90.052004}},
  \href{http://www.arXiv.org/abs/1406.3827}{\texttt{arXiv:1406.3827}}.

\bibitem{HIG-14-009}
\hrefCMSnoop {}{{CMS Collaboration}, ``{{Precise determination of the mass of
  the Higgs boson and tests of compatibility of its couplings with the standard
  model predictions using proton collisions at 7 and 8\TeV}}'',} \textit{ Eur.
  Phys. J. C} \textbf{ 75} (2015) 212,
  \href{http://dx.doi.org/10.1140/epjc/s10052-015-3351-7}{\doi{10.1140/epjc/s10052-015-3351-7}},
\href{http://www.arXiv.org/abs/1412.8662}{\texttt{arXiv:1412.8662}}.

\bibitem{Wess:1974tw}
\hrefCMSnoop {}{J.~Wess and B.~Zumino, ``{Supergauge transformations in four
  dimensions}'',} \textit{ Nucl. Phys. B} \textbf{ 70} (1974) 39,
\href{http://dx.doi.org/10.1016/0550-3213(74)90355-1}{\doi{10.1016/0550-3213(74)90355-1}}.

\bibitem{Cheng:2003ju}
\hrefCMSnoop {}{H.-C. Cheng and I.~Low, ``{TeV symmetry and the little
  hierarchy problem}'',} \textit{ JHEP} \textbf{ 09} (2003) 051,
  \href{http://dx.doi.org/10.1088/1126-6708/2003/09/051}{\doi{10.1088/1126-6708/2003/09/051}},
\href{http://www.arXiv.org/abs/hep-ph/0308199}{\texttt{arXiv:hep-ph/0308199}}.

\bibitem{Appelquist:2000nn}
\hrefCMSnoop {}{T.~Appelquist, H.-C. Cheng, and B.~A. Dobrescu, ``{Bounds on
  universal extra dimensions}'',} \textit{ Phys. Rev. D} \textbf{ 64} (2001)
  035002,
  \href{http://dx.doi.org/10.1103/PhysRevD.64.035002}{\doi{10.1103/PhysRevD.64.035002}},
\href{http://www.arXiv.org/abs/hep-ph/0012100}{\texttt{arXiv:hep-ph/0012100}}.

\bibitem{DarkMatt}
\hrefCMSnoop {}{G.~Bertone, D.~Hooper, and J.~Silk, ``{Particle dark matter:
  evidence, candidates and constraints}'',} \textit{ Phys. Rept.} \textbf{ 405}
  (2005) 279,
  \href{http://dx.doi.org/10.1016/j.physrep.2004.08.031}{\doi{10.1016/j.physrep.2004.08.031}},
\href{http://www.arXiv.org/abs/hep-ph/0404175}{\texttt{arXiv:hep-ph/0404175}}.

\bibitem{2HDM1}
\hrefCMSnoop {}{T.~D. Lee, ``A theory of spontaneous {$T$} violation'',}
  \textit{ Phys. Rev. D} \textbf{ 8} (1973) 1226,
\href{http://dx.doi.org/10.1103/PhysRevD.8.1226}{\doi{10.1103/PhysRevD.8.1226}}.

\bibitem{2HDM2}
\hrefCMSnoop {}{N.~G. Deshpande and E.~Ma, ``Pattern of symmetry breaking with
  two {Higgs} doublets'',} \textit{ Phys. Rev. D} \textbf{ 18} (1978) 2574,
\href{http://dx.doi.org/10.1103/PhysRevD.18.2574}{\doi{10.1103/PhysRevD.18.2574}}.

\bibitem{2HDM3}
\hrefCMSnoop {}{H.~E. Haber, G.~L. Kane, and T.~Sterling, ``The fermion mass
  scale and possible effects of {Higgs} bosons on experimental observables'',}
  \textit{ Nucl. Phys. B} \textbf{ 161} (1979) 493,
  \href{http://dx.doi.org/10.1016/0550-3213(79)90225-6}{\doi{10.1016/0550-3213(79)90225-6}}.

\bibitem{2HDM4}
G.~C. Branco\hrefCMSnoop {}{ {et~al.}, ``{Theory and phenomenology of
  two-Higgs-doublet models}'',} \textit{ Phys. Rept.} \textbf{ 516} (2012) 1,
  \href{http://dx.doi.org/10.1016/j.physrep.2012.02.002}{\doi{10.1016/j.physrep.2012.02.002}},
\href{http://www.arXiv.org/abs/1106.0034}{\texttt{arXiv:1106.0034}}.

\bibitem{2HDM5}
J.~F. Gunion, H.~E. Haber, G.~L. Kane, and S.~Dawson, ``The Higgs hunter's
  guide'', volume~80 of \textit{ Frontiers in Physics}.
\newblock Perseus Books, 2000.

\bibitem{pseudoscalar_8TeV}
\hrefCMSnoop {}{{CMS Collaboration}, ``{Search for a low-mass pseudoscalar
  Higgs boson produced in association with a $b\bar{b}$ pair in $pp$ collisions
  at $\sqrt{s} =$ 8 TeV}'',} \textit{ Phys. Lett. B} \textbf{ 758} (2016) 296,
  \href{http://dx.doi.org/10.1016/j.physletb.2016.05.003}{\doi{10.1016/j.physletb.2016.05.003}},
\href{http://www.arXiv.org/abs/1511.03610}{\texttt{arXiv:1511.03610}}.

\bibitem{pseudoscalar_8TeV_all}
\hrefCMSnoop {}{{CMS Collaboration}, ``{Search for light bosons in decays of
  the 125 GeV Higgs boson in proton-proton collisions at $ \sqrt{s}=8 $
  TeV}'',} \textit{ JHEP} \textbf{ 10} (2017) 076,
  \href{http://dx.doi.org/10.1007/JHEP10(2017)076}{\doi{10.1007/JHEP10(2017)076}},
\href{http://www.arXiv.org/abs/1701.02032}{\texttt{arXiv:1701.02032}}.

\bibitem{pseudoscalar_8TeV_ATLAS_mumutautau}
\hrefCMSnoop {}{{ATLAS Collaboration}, ``{Search for Higgs bosons decaying to
  $aa$ in the $\mu\mu\tau\tau$ final state in $pp$ collisions at $\sqrt{s} = $
  8 TeV with the ATLAS experiment}'',} \textit{ Phys. Rev. D} \textbf{ 92}
  (2015) 052002,
  \href{http://dx.doi.org/10.1103/PhysRevD.92.052002}{\doi{10.1103/PhysRevD.92.052002}},
\href{http://www.arXiv.org/abs/1505.01609}{\texttt{arXiv:1505.01609}}.

\bibitem{pseudoscalar_13TeV_mumutautau}
\hrefCMSnoop {}{{CMS Collaboration}, ``{Search for an exotic decay of the Higgs
  boson to a pair of light pseudoscalars in the final state of two muons and
  two $\tau$ leptons in proton-proton collisions at $ \sqrt{s}=13 $ TeV}'',}
  \textit{ JHEP} \textbf{ 11} (2018) 018,
  \href{http://dx.doi.org/10.1007/JHEP11(2018)018}{\doi{10.1007/JHEP11(2018)018}},
\href{http://www.arXiv.org/abs/1805.04865}{\texttt{arXiv:1805.04865}}.

\bibitem{pseudoscalar_13TeV}
\hrefCMSnoop {}{{CMS Collaboration}, ``{Search for an exotic decay of the Higgs
  boson to a pair of light pseudoscalars in the final state with two b quarks
  and two $\tau$ leptons in proton-proton collisions at $\sqrt{s}=$ 13 TeV}'',}
  \textit{ Phys. Lett. B} \textbf{ 785} (2018) 462,
  \href{http://dx.doi.org/10.1016/j.physletb.2018.08.057}{\doi{10.1016/j.physletb.2018.08.057}},
\href{http://www.arXiv.org/abs/1805.10191}{\texttt{arXiv:1805.10191}}.

\bibitem{pseudoscalar_8TeV_CMS_bbmumu}
\hrefCMSnoop {}{{CMS Collaboration}, ``{Search for a light pseudoscalar Higgs
  boson produced in association with bottom quarks in pp collisions at
  $\sqrt{s}=8 $ TeV}'',} \textit{ JHEP} \textbf{ 11} (2017)
  \href{http://dx.doi.org/10.1007/JHEP11(2017)010}{\doi{10.1007/JHEP11(2017)010}},
\href{http://www.arXiv.org/abs/1707.07283}{\texttt{arXiv:1707.07283}}.

\bibitem{pseudoscalar_13TeV_ATLAS_bbmumu}
\hrefCMSnoop {}{{ATLAS Collaboration}, ``{Search for Higgs boson decays into a
  pair of light bosons in the $bb\mu\mu$ final state in $pp$ collision at
  $\sqrt{s} = $13 TeV with the ATLAS detector}'',} \textit{ Phys. Lett. B}
  \textbf{ 790} (2019) 1,
  \href{http://dx.doi.org/10.1016/j.physletb.2018.10.073}{\doi{10.1016/j.physletb.2018.10.073}},
\href{http://www.arXiv.org/abs/1807.00539}{\texttt{arXiv:1807.00539}}.

\bibitem{pseudoscalar_13TeV_ATLAS_4b}
\hrefCMSnoop {}{{ATLAS Collaboration}, ``{Search for the Higgs boson produced
  in association with a vector boson and decaying into two spin-zero particles
  in the $H \rightarrow aa \rightarrow 4b$ channel in $pp$ collisions at
  $\sqrt{s} = 13$ TeV with the ATLAS detector}'',} \textit{ JHEP} \textbf{ 10}
  (2018) 031,
  \href{http://dx.doi.org/10.1007/JHEP10(2018)031}{\doi{10.1007/JHEP10(2018)031}},
\href{http://www.arXiv.org/abs/1806.07355}{\texttt{arXiv:1806.07355}}.

\bibitem{pseudoscalar_8TeV_ATLAS_4l}
\hrefCMSnoop {}{{ATLAS Collaboration}, ``{Search for new light gauge bosons in
  Higgs boson decays to four-lepton final states in $pp$ collisions at
  $\sqrt{s}=8$ TeV with the ATLAS detector at the LHC}'',} \textit{ Phys. Rev.
  D} \textbf{ 92} (2015) 092001,
  \href{http://dx.doi.org/10.1103/PhysRevD.92.092001}{\doi{10.1103/PhysRevD.92.092001}},
\href{http://www.arXiv.org/abs/1505.07645}{\texttt{arXiv:1505.07645}}.

\bibitem{pseudoscalar_13TeV_ATLAS_4l}
\hrefCMSnoop {}{{ATLAS Collaboration}, ``{Search for Higgs boson decays to
  beyond-the-Standard-Model light bosons in four-lepton events with the ATLAS
  detector at $\sqrt{s}=13$ TeV}'',} \textit{ JHEP} \textbf{ 06} (2018) 166,
  \href{http://dx.doi.org/10.1007/JHEP06(2018)166}{\doi{10.1007/JHEP06(2018)166}},
\href{http://www.arXiv.org/abs/1802.03388}{\texttt{arXiv:1802.03388}}.

\bibitem{pseudoscalar_8TeV_ATLAS_4g}
\hrefCMSnoop {}{{ATLAS Collaboration}, ``{Search for new phenomena in events
  with at least three photons collected in $pp$ collisions at $\sqrt{s}$ = 8
  TeV with the ATLAS detector}'',} \textit{ Eur. Phys. J. C} \textbf{ 76}
  (2016) 210,
  \href{http://dx.doi.org/10.1140/epjc/s10052-016-4034-8}{\doi{10.1140/epjc/s10052-016-4034-8}},
\href{http://www.arXiv.org/abs/1509.05051}{\texttt{arXiv:1509.05051}}.

\bibitem{VLQ1}
\hrefCMSnoop {}{J.~A. Aguilar-Saavedra, R.~Benbrik, S.~Heinemeyer, and
  M.~P{\'e}rez-Victoria, ``Handbook of vectorlike quarks: Mixing and single
  production'',} \textit{ Phys. Rev. D} \textbf{ 88} (2013) 094010,
  \href{http://dx.doi.org/10.1103/PhysRevD.88.094010}{\doi{10.1103/PhysRevD.88.094010}},
\href{http://www.arXiv.org/abs/1306.0572}{\texttt{arXiv:1306.0572}}.

\bibitem{VLQ2}
\hrefCMSnoop {}{P.~W. Graham, A.~Ismail, S.~Rajendran, and P.~Saraswat,
  ``{Little solution to the little hierarchy problem: A vectorlike
  generation}'',} \textit{ Phys. Rev. D} \textbf{ 81} (2010) 055016,
  \href{http://dx.doi.org/10.1103/PhysRevD.81.055016}{\doi{10.1103/PhysRevD.81.055016}},
\href{http://www.arXiv.org/abs/0910.3020}{\texttt{arXiv:0910.3020}}.

\bibitem{VLQ_LHC1}
\hrefCMSnoop {}{A.~Angelescu, A.~Djouadi, and G.~Moreau, ``{Vector-like
  top/bottom quark partners and Higgs physics at the LHC}'',} \textit{ Eur.
  Phys. J. C} \textbf{ 76} (2016) 99,
  \href{http://dx.doi.org/10.1140/epjc/s10052-016-3950-y}{\doi{10.1140/epjc/s10052-016-3950-y}},
\href{http://www.arXiv.org/abs/1510.07527}{\texttt{arXiv:1510.07527}}.

\bibitem{VLQ_LHC2}
\hrefCMSnoop {}{Y.~Okada and L.~Panizzi, ``{LHC signatures of vector-like
  quarks}'',} \textit{ Adv. High Energy Phys.} \textbf{ 2013} (2013) 364936,
  \href{http://dx.doi.org/10.1155/2013/364936}{\doi{10.1155/2013/364936}},
\href{http://www.arXiv.org/abs/1207.5607}{\texttt{arXiv:1207.5607}}.

\bibitem{HVM1}
\hrefCMSnoop {}{J.~S. Matthew and K.~M. Zurek, ``Echoes of a hidden valley at
  hadron colliders'',} \textit{ Phys. Lett. B} \textbf{ 651} (2007) 374,
  \href{http://dx.doi.org/10.1016/j.physletb.2007.06.055}{\doi{10.1016/j.physletb.2007.06.055}},
\href{http://www.arXiv.org/abs/hep-ph/0604261}{\texttt{arXiv:hep-ph/0604261}}.

\bibitem{HVM2}
\hrefCMSnoop {}{T.~Han, Z.~Si, K.~M. Zurek, and M.~J. Strassler,
  ``Phenomenology of hidden valleys at hadron colliders'',} \textit{ JHEP}
  \textbf{ 07} (2008) 008,
  \href{http://dx.doi.org/10.1088/1126-6708/2008/07/008}{\doi{10.1088/1126-6708/2008/07/008}},
\href{http://www.arXiv.org/abs/0712.2041}{\texttt{arXiv:0712.2041}}.

\bibitem{DFM1}
\hrefCMSnoop {}{K.~Kong, H.~S. Lee, and M.~Park, ``Dark decay of the top
  quark'',} \textit{ Phys. Rev. D} \textbf{ 89} (2014) 074007,
  \href{http://dx.doi.org/10.1103/PhysRevD.89.074007}{\doi{10.1103/PhysRevD.89.074007}},
\href{http://www.arXiv.org/abs/1401.5020}{\texttt{arXiv:1401.5020}}.

\bibitem{lowmassZ2}
\hrefCMSnoop {}{W.~Altmannshofer, S.~Gori, S.~Profumo, and F.~S. Queiroz,
  ``Explaining dark matter and b decay anomalies with an $l_\mu - l_\tau$
  model'',} \textit{ JHEP} \textbf{ 12} (2016) 106,
  \href{http://dx.doi.org/10.1007/JHEP12(2016)106}{\doi{10.1007/JHEP12(2016)106}},
\href{http://www.arXiv.org/abs/1609.04026}{\texttt{arXiv:1609.04026}}.

\bibitem{Dalchenko:2017shg}
M.~Abdullah\hrefCMSnoop {}{ {et~al.}, ``{Bottom-quark fusion processes at the
  LHC for probing Z$^{'}$ models and $B$-meson decay anomalies}'',} \textit{
  Phys. Rev. D} \textbf{ 97} (2018) 075035,
  \href{http://dx.doi.org/10.1103/PhysRevD.97.075035}{\doi{10.1103/PhysRevD.97.075035}},
\href{http://www.arXiv.org/abs/1707.07016}{\texttt{arXiv:1707.07016}}.

\bibitem{FCNC1}
\hrefCMSnoop {}{S.~Jung, H.~Murayama, A.~Pierce, and J.~D. Wells, ``Top quark
  forward-backward asymmetry from new $t$-channel physics'',} \textit{ Phys.
  Rev. D} \textbf{ 81} (2010) 015004,
  \href{http://dx.doi.org/10.1103/PhysRevD.81.015004}{\doi{10.1103/PhysRevD.81.015004}},
\href{http://www.arXiv.org/abs/0907.4112}{\texttt{arXiv:0907.4112}}.

\bibitem{dimuon_CMS}
\hrefCMSnoop {}{{CMS Collaboration}, ``{Search for resonances in the mass
  spectrum of muon pairs produced in association with b quark jets in
  proton-proton collisions at $\sqrt{s} =$ 8 and 13 TeV}'',} \textit{ JHEP}
  \textbf{ 11} (2018) 161,
  \href{http://dx.doi.org/10.1007/JHEP11(2018)161}{\doi{10.1007/JHEP11(2018)161}},
\href{http://www.arXiv.org/abs/1808.01890}{\texttt{arXiv:1808.01890}}.

\bibitem{trigger}
\hrefCMSnoop {}{{CMS Collaboration}, ``{The CMS trigger system}'',} \textit{
  JINST} \textbf{ 12} (2017) P01020,
  \href{http://dx.doi.org/10.1088/1748-0221/12/01/P01020}{\doi{10.1088/1748-0221/12/01/P01020}},
\href{http://www.arXiv.org/abs/1609.02366}{\texttt{arXiv:1609.02366}}.

\bibitem{CMS}
\hrefCMSnoop {}{{CMS Collaboration}, ``{The CMS experiment at the CERN LHC}'',}
  \textit{ JINST} \textbf{ 3} (2008) S08004,
  \href{http://dx.doi.org/10.1088/1748-0221/3/08/S08004}{\doi{10.1088/1748-0221/3/08/S08004}},
\href{http://www.arXiv.org/abs/1510.07488}{\texttt{arXiv:1510.07488}}.

\bibitem{aMCatNLO}
J.~Alwall\hrefCMSnoop {}{ {et~al.}, ``{The automated computation of tree-level
  and next-to-leading order differential cross sections, and their matching to
  parton shower simulations}'',} \textit{ JHEP} \textbf{ 07} (2014) 079,
  \href{http://dx.doi.org/10.1007/JHEP07(2014)079}{\doi{10.1007/JHEP07(2014)079}},
  \href{http://www.arXiv.org/abs/1405.0301}{\texttt{arXiv:1405.0301}}.

\bibitem{MLM}
J.~Alwall\hrefCMSnoop {}{ {et~al.}, ``Comparative study of various algorithms
  for the merging of parton showers and matrix elements in hadronic
  collisions'',} \textit{ Eur. Phys. J. C} \textbf{ 53} (2008) 473,
  \href{http://dx.doi.org/10.1140/epjc/s10052-007-0490-5}{\doi{10.1140/epjc/s10052-007-0490-5}},
  \href{http://www.arXiv.org/abs/0706.2569}{\texttt{arXiv:0706.2569}}.

\bibitem{FxFx}
\hrefCMSnoop {}{R.~Frederix and S.~Frixione, ``{Merging meets matching in
  MC@NLO}'',} \textit{ JHEP} \textbf{ 12} (2012) 061,
  \href{http://dx.doi.org/10.1007/JHEP12(2012)061}{\doi{10.1007/JHEP12(2012)061}},
  \href{http://www.arXiv.org/abs/1209.6215}{\texttt{arXiv:1209.6215}}.

\bibitem{POWHEG0}
\hrefCMSnoop {}{P.~Nason, ``{A new method for combining NLO QCD with shower
  Monte Carlo algorithms}'',} \textit{ JHEP} \textbf{ 11} (2004) 040,
  \href{http://dx.doi.org/10.1088/1126-6708/2004/11/040}{\doi{10.1088/1126-6708/2004/11/040}},
  \href{http://www.arXiv.org/abs/hep-ph/0409146}{\texttt{arXiv:hep-ph/0409146}}.

\bibitem{POWHEG1}
\hrefCMSnoop {}{S.~Frixione, P.~Nason, and C.~Oleari, ``{Matching NLO QCD
  computations with parton shower simulations: the POWHEG method}'',} \textit{
  JHEP} \textbf{ 11} (2007) 070,
  \href{http://dx.doi.org/10.1088/1126-6708/2007/11/070}{\doi{10.1088/1126-6708/2007/11/070}},
  \href{http://www.arXiv.org/abs/0709.2092}{\texttt{arXiv:0709.2092}}.

\bibitem{POWHEG2}
\hrefCMSnoop {}{S.~Alioli, P.~Nason, C.~Oleari, and E.~Re, ``{A general
  framework for implementing {NLO} calculations in shower Monte Carlo programs:
  the POWHEG BOX}'',} \textit{ JHEP} \textbf{ 06} (2010) 043,
  \href{http://dx.doi.org/10.1007/JHEP06(2010)043}{\doi{10.1007/JHEP06(2010)043}},
  \href{http://www.arXiv.org/abs/1002.2581}{\texttt{arXiv:1002.2581}}.

\bibitem{POWHEG_TT}
\hrefCMSnoop {}{S.~Frixione, P.~Nason, and G.~Ridolfi, ``{A positive-weight
  next-to-leading-order Monte Carlo for heavy flavour hadroproduction}'',}
  \textit{ JHEP} \textbf{ 09} (2007) 126,
  \href{http://dx.doi.org/10.1088/1126-6708/2007/09/126}{\doi{10.1088/1126-6708/2007/09/126}},
\href{http://www.arXiv.org/abs/0707.3088}{\texttt{arXiv:0707.3088}}.

\bibitem{POWHEG_TT2}
\hrefCMSnoop {}{J.~M. Campbell, R.~K. Ellis, P.~Nason, and E.~Re, ``Top-pair
  production and decay at {NLO} matched with parton showers'',} \textit{ JHEP}
  \textbf{ 04} (2015) 114,
  \href{http://dx.doi.org/10.1007/JHEP04(2015)114}{\doi{10.1007/JHEP04(2015)114}},
\href{http://www.arXiv.org/abs/1412.1828}{\texttt{arXiv:1412.1828}}.

\bibitem{POWHEG_ST_st}
\hrefCMSnoop {}{S.~Alioli, P.~Nason, C.~Oleari, and E.~Re, ``{NLO single-top
  production matched with shower in POWHEG: $s$- and $t$-channel
  contributions}'',} \textit{ JHEP} \textbf{ 09} (2009) 111,
  \href{http://dx.doi.org/10.1088/1126-6708/2009/09/111}{\doi{10.1088/1126-6708/2009/09/111}},
  \href{http://www.arXiv.org/abs/0907.4076}{\texttt{arXiv:0907.4076}}.
  [Erratum: \DOI{10.1007/JHEP02(2010)011}].

\bibitem{POWHEG_ST_tW}
\hrefCMSnoop {}{E.~Re, ``{Single-top Wt-channel production matched with parton
  showers using the POWHEG method}'',} \textit{ Eur. Phys. J. C} \textbf{ 71}
  (2011) 1547,
  \href{http://dx.doi.org/10.1140/epjc/s10052-011-1547-z}{\doi{10.1140/epjc/s10052-011-1547-z}},
  \href{http://www.arXiv.org/abs/1009.2450}{\texttt{arXiv:1009.2450}}.

\bibitem{sigma_DY}
\hrefCMSnoop {}{Y.~Li and F.~Petriello, ``{Combining QCD and electroweak
  corrections to production in FEWZ}'',} \textit{ Phys. Rev. D} \textbf{ 86}
  (2012) 094034,
  \href{http://dx.doi.org/10.1103/PhysRevD.86.094034}{\doi{10.1103/PhysRevD.86.094034}},
\href{http://www.arXiv.org/abs/1208.5967}{\texttt{arXiv:1208.5967}}.

\bibitem{sigma_TT}
\hrefCMSnoop {}{M.~Czakon and A.~Mitov, ``Top++: A program for the calculation
  of the top-pair cross-section at hadron colliders'',} \textit{ Comput. Phys.
  Commun.} \textbf{ 185} (2014) 2930,
  \href{http://dx.doi.org/10.1016/j.cpc.2014.06.021}{\doi{10.1016/j.cpc.2014.06.021}},
  \href{http://www.arXiv.org/abs/1112.5675}{\texttt{arXiv:1112.5675}}.

\bibitem{sigma_ST}
P.~Kant\hrefCMSnoop {}{ {et~al.}, ``{HatHor for single top-quark production:
  Updated predictions and uncertainty estimates for single top-quark production
  in hadronic collisions}'',} \textit{ Comput. Phys. Commun.} \textbf{ 191}
  (2015) 74,
  \href{http://dx.doi.org/10.1016/j.cpc.2015.02.001}{\doi{10.1016/j.cpc.2015.02.001}},
\href{http://www.arXiv.org/abs/1406.4403}{\texttt{arXiv:1406.4403}}.

\bibitem{Pythia8_2015}
T.~Sj{\"o}strand\hrefCMSnoop {}{ {et~al.}, ``An introduction to {PYTHIA
  8.2}'',} \textit{ Comput. Phys. Commun.} \textbf{ 191} (2015) 159,
  \href{http://dx.doi.org/10.1016/j.cpc.2015.01.024}{\doi{10.1016/j.cpc.2015.01.024}},
  \href{http://www.arXiv.org/abs/1410.3012}{\texttt{arXiv:1410.3012}}.

\bibitem{CMS-TOP-17-011}
\hrefCMSnoop {}{{CMS Collaboration}, ``{Measurement of the single top quark and
  antiquark production cross sections in the $t$ channel and their ratio in
  proton-proton collisions at $\sqrt{s}=$ 13 TeV}'',} (2018).
  \href{http://www.arXiv.org/abs/1812.10514}{\texttt{arXiv:1812.10514}}.
Submitted to \textit{Phys. Lett. B.}

\bibitem{CUETP8M1}
\hrefCMSnoop {}{{CMS Collaboration}, ``{Event generator tunes obtained from
  underlying event and multiparton scattering measurements}'',} \textit{ Eur.
  Phys. J. C} \textbf{ 76} (2016) 155,
  \href{http://dx.doi.org/10.1140/epjc/s10052-016-3988-x}{\doi{10.1140/epjc/s10052-016-3988-x}},
  \href{http://www.arXiv.org/abs/1512.00815}{\texttt{arXiv:1512.00815}}.

\bibitem{NNPDF30}
R.~D. Ball\hrefCMSnoop {}{ {et~al.}, ``{Parton distributions for the LHC Run
  II}'',} \textit{ JHEP} \textbf{ 15} (2015) 40,
  \href{http://dx.doi.org/10.1007/JHEP04(2015)040}{\doi{10.1007/JHEP04(2015)040}},
\href{http://www.arXiv.org/abs/1410.8849}{\texttt{arXiv:1410.8849}}.

\bibitem{GEANT4}
\hrefCMSnoop {}{{GEANT4} Collaboration, ``{GEANT4} --- a simulation toolkit'',}
  \textit{ Nucl. Instrum. Meth. A} \textbf{ 506} (2003) 250,
\href{http://dx.doi.org/10.1016/S0168-9002(03)01368-8}{\doi{10.1016/S0168-9002(03)01368-8}}.

\bibitem{CMS-PAS-LUM-17-001}
\href {https://cds.cern.ch/record/2257069}{{CMS Collaboration}, ``{CMS}
  luminosity measurements for the 2016 data taking period'',} CMS Physics
  Analysis Summary CMS-PAS-LUM-17-001, CERN, 2017.

\bibitem{PF2017}
\hrefCMSnoop {}{{CMS Collaboration}, ``{Particle-flow reconstruction and global
  event description with the CMS detector}'',} \textit{ JINST} \textbf{ 12}
  (2017) P10003,
  \href{http://dx.doi.org/10.1088/1748-0221/12/10/P10003}{\doi{10.1088/1748-0221/12/10/P10003}},
\href{http://www.arXiv.org/abs/1706.04965}{\texttt{arXiv:1706.04965}}.

\bibitem{TMVA}
\href {http://pos.sissa.it/archive/conferences/050/040/ACAT_040.pdf}{H.~Voss,
  A.~H{\"o}cker, J.~Stelzer, and F.~Tegenfeldt, ``{TMVA}, the toolkit for
  multivariate data analysis with {ROOT}'',} in \textit{ {XI Int. Workshop on
  Advanced Computing and Analysis Techniques in Physics Research}}.
\newblock 2007.
\newblock
  \href{http://www.arXiv.org/abs/physics/0703039}{\texttt{arXiv:physics/0703039}}.
\newblock
{PoS ACAT:040}.

\bibitem{electron}
\hrefCMSnoop {}{{CMS Collaboration}, ``{Performance of electron reconstruction
  and selection with the CMS detector in proton-proton collisions at $\sqrt{s}
  = 8\,\text{TeV}$}'',} \textit{ JINST} \textbf{ 10} (2015) P06005,
  \href{http://dx.doi.org/10.1088/1748-0221/10/06/P06005}{\doi{10.1088/1748-0221/10/06/P06005}},
\href{http://www.arXiv.org/abs/1502.02701}{\texttt{arXiv:1502.02701}}.

\bibitem{Sirunyan:2018fpa}
\hrefCMSnoop {}{{CMS Collaboration}, ``{Performance of the CMS muon detector
  and muon reconstruction with proton-proton collisions at $\sqrt{s}=$ 13
  TeV}'',} \textit{ JINST} \textbf{ 13} (2018) P06015,
  \href{http://dx.doi.org/10.1088/1748-0221/13/06/P06015}{\doi{10.1088/1748-0221/13/06/P06015}},
\href{http://www.arXiv.org/abs/1804.04528}{\texttt{arXiv:1804.04528}}.

\bibitem{antikT}
\hrefCMSnoop {}{M.~Cacciari, G.~P. Salam, and G.~Soyez, ``The anti-\kt jet
  clustering algorithm'',} \textit{ JHEP} \textbf{ 04} (2008) 063,
  \href{http://dx.doi.org/10.1088/1126-6708/2008/04/063}{\doi{10.1088/1126-6708/2008/04/063}},
\href{http://www.arXiv.org/abs/0802.1189}{\texttt{arXiv:0802.1189}}.

\bibitem{fastjet1}
\hrefCMSnoop {}{M.~Cacciari, G.~P. Salam, and G.~Soyez, ``{FastJet user
  manual}'',} \textit{ Eur. Phys. J. C} \textbf{ 72} (2012) 1896,
  \href{http://dx.doi.org/10.1140/epjc/s10052-012-1896-2}{\doi{10.1140/epjc/s10052-012-1896-2}},
\href{http://www.arXiv.org/abs/1111.6097}{\texttt{arXiv:1111.6097}}.

\bibitem{fastjet2}
\hrefCMSnoop {}{{M. Cacciari and G. P. Salam}, ``{Dispelling the $N^{3}$ myth
  for the $\kt$ jet-finder}'',} \textit{ Phys. Lett. B} \textbf{ 641} (2006)
  57,
  \href{http://dx.doi.org/10.1016/j.physletb.2006.08.037}{\doi{10.1016/j.physletb.2006.08.037}},
\href{http://www.arXiv.org/abs/hep-ph/0512210}{\texttt{arXiv:hep-ph/0512210}}.

\bibitem{CMS-JME-10-011}
\hrefCMSnoop {}{{CMS Collaboration}, ``{Determination of jet energy calibration
  and transverse momentum resolution in CMS}'',} \textit{ JINST} \textbf{ 6}
  (2011) P11002,
  \href{http://dx.doi.org/10.1088/1748-0221/6/11/P11002}{\doi{10.1088/1748-0221/6/11/P11002}},
\href{http://www.arXiv.org/abs/1107.4277}{\texttt{arXiv:1107.4277}}.

\bibitem{jetPUID}
\href {http://cdsweb.cern.ch/record/1581583}{{CMS Collaboration}, ``Pileup jet
  identification'',} {CMS Physics Analysis Summary} CMS-PAS-JME-13-005, CERN,
  2013.

\bibitem{jetID_2016}
\href {https://cds.cern.ch/record/2256875}{{CMS Collaboration}, ``{Jet
  algorithms performance in 13 TeV data}'',} {CMS Physics Analysis Summary}
  CMS-PAS-JME-16-003, CERN, 2017.

\bibitem{btagging}
\hrefCMSnoop {}{{CMS Collaboration}, ``{Identification of heavy-flavour jets
  with the CMS detector in pp collisions at 13 TeV}'',} \textit{ JINST}
  \textbf{ 13} (2018) P05011,
  \href{http://dx.doi.org/10.1088/1748-0221/13/05/P05011}{\doi{10.1088/1748-0221/13/05/P05011}},
\href{http://www.arXiv.org/abs/1712.07158}{\texttt{arXiv:1712.07158}}.

\bibitem{HPS2}
\hrefCMSnoop {}{{CMS Collaboration}, ``{Performance of reconstruction and
  identification of $\tau$ leptons decaying to hadrons and $\nu_\tau$ in pp
  collisions at $\sqrt{s}=$ 13 TeV}'',} \textit{ JINST} \textbf{ 13} (2018)
  P10005,
  \href{http://dx.doi.org/10.1088/1748-0221/13/10/P10005}{\doi{10.1088/1748-0221/13/10/P10005}},
\href{http://www.arXiv.org/abs/1809.02816}{\texttt{arXiv:1809.02816}}.

\bibitem{CMS-PAS-JME-16-004}
\href {http://cds.cern.ch/record/2205284}{{CMS Collaboration}, ``{Performance
  of missing energy reconstruction in 13 TeV pp collision data using the CMS
  detector}'',} {CMS Physics Analysis Summary} CMS-PAS-JME-16-004, CERN, 2016.

\bibitem{CMS-PAS-JME-17-001}
\href {http://cds.cern.ch/record/2628600}{{CMS Collaboration}, ``{Performance
  of missing transverse momentum in pp collisions at $\sqrt{s}=13$~TeV using
  the CMS detector}'',} {CMS Physics Analysis Summary} CMS-PAS-JME-17-001,
  CERN, 2018.

\bibitem{pzeta_CDF}
\hrefCMSnoop {}{{CDF} Collaboration, ``{Search for neutral MSSM Higgs bosons
  decaying to tau pairs in $p\bar{p}$ collisions at $\sqrt{s} = 1.96$ TeV}'',}
  \textit{ Phys. Rev. Lett.} \textbf{ 96} (2006) 011802,
  \href{http://dx.doi.org/10.1103/PhysRevLett.96.011802}{\doi{10.1103/PhysRevLett.96.011802}},
\href{http://www.arXiv.org/abs/hep-ex/0508051}{\texttt{arXiv:hep-ex/0508051}}.

\bibitem{SVFit2014}
\hrefCMSnoop {}{L.~Bianchini, J.~Conway, E.~K. Friis, and C.~Veelken,
  ``Reconstruction of the {Higgs} mass in {$H \to \tau\tau$} events by
  dynamical likelihood techniques'',} \textit{ J. Phys. Conf. Ser.} \textbf{
  513} (2014) 022035,
\href{http://dx.doi.org/10.1088/1742-6596/513/2/022035}{\doi{10.1088/1742-6596/513/2/022035}}.

\bibitem{SVFit2017}
L.~Bianchini\hrefCMSnoop {}{ {et~al.}, ``{Reconstruction of the Higgs mass in
  events with Higgs bosons decaying into a pair of $\tau$ leptons using matrix
  element techniques}'',} \textit{ Nucl. Instrum. Meth. A} \textbf{ 862} (2017)
  54,
  \href{http://dx.doi.org/10.1016/j.nima.2017.05.001}{\doi{10.1016/j.nima.2017.05.001}},
\href{http://www.arXiv.org/abs/1603.05910}{\texttt{arXiv:1603.05910}}.

\bibitem{LHC-HCG-Report}
\href {https://cds.cern.ch/record/1379837}{{ATLAS and CMS Collaborations},
  ``Procedure for the {LHC} {Higgs} boson search combination in summer 2011'',}
  Technical Report CMS-NOTE-2011-005, ATL-PHYS-PUB-2011-011, CERN, 2011.

\bibitem{HTT_2016_paper}
\hrefCMSnoop {}{{CMS Collaboration}, ``{Observation of the Higgs boson decay to
  a pair of $\tau$ leptons with the CMS detector}'',} \textit{ Phys. Lett. B}
  \textbf{ 779} (2018) 283,
  \href{http://dx.doi.org/10.1016/j.physletb.2018.02.004}{\doi{10.1016/j.physletb.2018.02.004}},
\href{http://www.arXiv.org/abs/1708.00373}{\texttt{arXiv:1708.00373}}.

\bibitem{HPS3}
\href {http://cds.cern.ch/record/2196972}{{CMS Collaboration}, ``{Performance
  of reconstruction and identification of tau leptons in their decays to
  hadrons and tau neutrino in LHC Run-2}'',} {CMS Physics Analysis Summary}
  CMS-PAS-TAU-16-002, CERN, 2016.

\bibitem{deFlorian:2016spz}
\hrefCMSnoop {}{D.~de~Florian {et~al.}, ``Handbook of {LHC} {H}iggs cross
  sections: 4. deciphering the nature of the {H}iggs sector'',} CERN Report
  CERN-2017-002-M, 2016.
\newblock
  \href{http://dx.doi.org/10.23731/CYRM-2017-002}{\doi{10.23731/CYRM-2017-002}},
  \href{http://www.arXiv.org/abs/1610.07922}{\texttt{arXiv:1610.07922}}.

\bibitem{CLs1}
\hrefCMSnoop {}{G.~Cowan, K.~Cranmer, E.~Gross, and O.~Vitells, ``Asymptotic
  formulae for likelihood-based tests of new physics'',} \textit{ Eur. Phys. J.
  C} \textbf{ 71} (2011)
  \href{http://dx.doi.org/10.1140/epjc/s10052-011-1554-0}{\doi{10.1140/epjc/s10052-011-1554-0}},
  \href{http://www.arXiv.org/abs/1007.1727}{\texttt{arXiv:1007.1727}}.
[Erratum: \DOI{10.1140/epjc/s10052-013-2501-z}].

\bibitem{Junk:1999kv}
\hrefCMSnoop {}{T.~Junk, ``{Confidence level computation for combining searches
  with small statistics}'',} \textit{ Nucl. Instrum. Meth. A} \textbf{ 434}
  (1999) 435,
  \href{http://dx.doi.org/10.1016/S0168-9002(99)00498-2}{\doi{10.1016/S0168-9002(99)00498-2}},
  \href{http://www.arXiv.org/abs/hep-ex/9902006}{\texttt{arXiv:hep-ex/9902006}}.

\bibitem{CLs3}
\hrefCMSnoop {}{A.~L. Read, ``{Presentation of search results: The
  CL$_\text{s}$ technique}'',} \textit{ J. Phys. G} \textbf{ 28} (2002) 2693,
\href{http://dx.doi.org/10.1088/0954-3899/28/10/313}{\doi{10.1088/0954-3899/28/10/313}}.

\bibitem{2HDM_8TeV}
\hrefCMSnoop {}{J.~Bernon, J.~F. Gunion, Y.~Jiang, and S.~Kraml, ``{Light Higgs
  bosons in Two-Higgs-Doublet Models}'',} \textit{ Phys. Rev. D} \textbf{ 19}
  (2015) 075019,
  \href{http://dx.doi.org/10.1103/PhysRevD.91.075019}{\doi{10.1103/PhysRevD.91.075019}},
\href{http://www.arXiv.org/abs/1412.3385}{\texttt{arXiv:1412.3385}}.

\end{thebibliography}\endgroup
